\documentclass[sigconf]{acmart}
\usepackage{graphicx}
\usepackage{xspace}
\usepackage{epsfig}
\usepackage{multirow}
\usepackage{url}
\usepackage{booktabs}
\usepackage{amsmath}
\usepackage{balance}
\usepackage{pifont}
\usepackage{rotating}

\begin{abstract}
	According to common relevance-judgments regimes, such as TREC's, a document can be
	deemed relevant to a query even if it contains a very short passage of
	text with pertinent information. This fact has motivated work on
	passage-based document retrieval: document ranking methods that induce
	information from the document's passages. However, the main source of passage-based information utilized was passage-query
	similarities. We address the challenge of utilizing richer sources of
	passage-based information to improve document retrieval
	effectiveness. Specifically, we devise a suite of learning-to-rank-based document
	retrieval methods that utilize an effective ranking of passages
	produced in response to the query; the passage ranking is also induced using a
	learning-to-rank approach. Some of the methods quantify the ranking of the passages of a document. Others utilize the feature-based representation of passages used for learning a passage ranker.
	Empirical evaluation attests to the clear
	merits of our methods with respect to highly effective baselines.
	Our best performing method is based on learning a document ranking function using document-query features and passage-query features of the document's passage most highly ranked.
\end{abstract}

\copyrightyear{2018}
\acmYear{}
\setcopyright{acmcopyright}
\acmConference[]{}{}{}
\acmBooktitle{}
\acmPrice{15.00}
\acmDOI{}
\acmISBN{}

\newcommand{\myparagraph}[1]{\vspace{0.3\baselineskip}\noindent{\textbf{#1}}.~}
\newcommand{\omt}[1]{}

\newcommand{\init}{init}

\newcommand{\doc}{d}

\newcommand{\corpus}{D}
\newcommand{\coll}{\corpus}

\newcommand{\query}{q\xspace}

\newcommand{\passage}{g}

\newcommand{\inex}{INEX\xspace}
\newcommand{\aquaint}{AQUAINT\xspace}
\newcommand{\inquery}{INQUERY\xspace}

\newcommand{\iP}{iP[x]\xspace}
\newcommand{\officialiPOne}{iP[.1]\xspace}
\newcommand{\officialiP}{iP[.01]\xspace}
\newcommand{\MAiP}{MAiP\xspace}
\newcommand{\MAP}{MAP\xspace}

\newcommand{\PTen}{p@10\xspace}

\newcommand{\psgLength}[1]{Psg#1}

\newcommand{\QuerySimFuse}{QSF\xspace}

\newcommand{\PBCShort}{PLM\xspace}

\newcommand{\PLMShort}{PLM\xspace}

\newcommand{\owpc}{owpc\xspace}

\newcommand{\genScoreList}[2]{\scoreWord_{#2}(#1)}
\newcommand{\arbList}{L}
\newcommand{\arbItem}{x}

\newcommand{\SVM}{RankSVM\xspace}
\newcommand{\SVMabb}{SVM\xspace}

\newcommand{\LambdaMART}{LambdaMART\xspace}
\newcommand{\LambdaMARTabb}{LMart\xspace}

\newcommand{\MART}{MART\xspace}

\newcommand{\CAscent}{CAscent\xspace}

\newcommand{\firstmention}[1]{{\bf #1}}

\newcommand{\scoreWord}{Score}

\newcommand{\passageCE}{PsgQuerySim\xspace}
\newcommand{\documentSim}{DocQuerySim\xspace}

\newcommand{\pAvg}{AvgPDSim\xspace}

\newcommand{\pMax}{MaxPDSim\xspace}

\newcommand{\lengthRatio}{LengthRatio\xspace}
\newcommand{\std}{StdPDSim\xspace}
\newcommand{\entropy}{Ent\xspace}
\newcommand{\nighborLeft}{QuerySimPre\xspace}
\newcommand{\nighborRight}{QuerySimFollow\xspace}
\newcommand{\queryTermCount}{QueryLength\xspace}
\newcommand{\swOne}{SW1\xspace}
\newcommand{\swTwo}{SW2\xspace}
\newcommand{\MKexactMatch}{ExactMatch\xspace}
\newcommand{\MKtermOverlap}{TermOverlap\xspace}
\newcommand{\MKsynonymsOverlap}{SynonymsOverlap\xspace}
\newcommand{\MKgLength}{PsgLength\xspace}
\newcommand{\MKgLocation}{PsgLocation\xspace}
\newcommand{\ESA}{ESA\xspace}

\newcommand{\SemWordEmbedding}{W2V\xspace}

\newcommand{\EntityLinking}{Entity\xspace}
\newcommand{\RFrac}{RFrac\xspace}

\newcommand{\MKS}{MKS\xspace}

\newcommand{\definedas}{\stackrel{def}{=}}

\newcommand{\sigDoc}{l}

\newcommand{\sigQSF}{f}

\newcommand{\sigOwpc}{o}
\newcommand{\sigPLM}{m}

\newcommand{\sigHome}{d}
\newcommand{\sigMRF}{s}
\newcommand{\sigMKS}{k}

\newcommand{\sigDocLTR}{i}

\newcommand{\sigRanking}{*}

\newcommand{\ce}[2]{CE(#1\, ||\, #2)}

\newcommand{\simFn}[2]{Sim(#1,#2)}

\newcommand{\topRetGroup}{\corpus_{init}}
\newcommand{\topRetLTRList}{\corpus_{LTR}}

\newcommand{\initPsgGroup}{G(\topRetLTRList)}
\newcommand{\topPsgLTRList}{\initPsgGroup}

\newcommand{\docOf}{\doc_{\passage}}

\newcommand{\neighborLeft}{QuerySimPre\xspace}
\newcommand{\neighborRight}{QuerySimFollow\xspace}

\newcommand{\QL}{LM\xspace}

\newcommand{\HLabb}{DocPsg\xspace}
\newcommand{\MRF}{MRF\xspace}

\newcommand{\RRFshort}{RRF\xspace}
\newcommand{\SDM}{SDM\xspace}

\newcommand{\Mslr}{MSLR\xspace}

\newcommand{\MethodTwoTitle}{SMPD\xspace}

\newcommand{\MethodThreeabb}{JPDs\xspace}
\newcommand{\JPDSecond}{JPDs-second\xspace}
\newcommand{\JPDThird}{JPDs-third\xspace}
\newcommand{\JPDLast}{JPDs-lowest\xspace}
\newcommand{\JPDm}{JPDm\xspace}
\newcommand{\JPDAvg}{JPDm-avg\xspace}
\newcommand{\JPDMax}{JPDm-max\xspace}
\newcommand{\JPDMin}{JPDm-min\xspace}

\newcommand{\fVector}[2]{\vec{v}_{(#1,#2)}}
\newcommand{\fVectorParam}[3]{\vec{v}^{#3}_{(#1,#2)}}
\newcommand{\fVectorTwo}[2]{\vec{v'}_{(#1,#2)}}

\newcommand{\rrfParam}{\nu}
\newcommand{\rankFunc}[2]{r_{#2}(#1)}

\newcommand{\docList}{S_{doc}}
\newcommand{\psgList}{S_{psg}}

\newcommand{\fpd}{FPD\xspace}

\newcommand{\sigBP}{j}

\newcommand{\psgLTR}{PsgLTR\xspace}
\newcommand{\psgLM}{QSF\xspace}

\newcommand{\initSet}{S^{init}_{psg}}

\begin{document}

\title{A Passage-Based Approach to Learning to Rank Documents}

\author{Eilon Sheetrit}
\email{seilon@campus.technion.ac.il}
\affiliation{Technion}

\author{Anna Shtok}
\email{annie.shtok@gmail.com}
\affiliation{%
}

\author{Oren Kurland}
\email{kurland@ie.technion.ac.il}
\affiliation{%
  \institution{Technion}
}

\renewcommand{\shortauthors}{Sheetrit et al.}

\maketitle

\section{Introduction}
\label{sec:intro}
The ad hoc retrieval task is ranking documents in a corpus in response
to a query by presumed relevance to the information need 
the query represents. Often, documents are deemed relevant even if
they contain only a short passage with pertaining
information; e.g., by TREC's relevance judgment regime
\citep{Voorhees+Harman:05a}.  

As a result, there has been a large body of work on {\em passage-based
	document retrieval}: utilizing information induced from
document passages to rank the documents; e.g.,  \cite{callan:1994,Wilkinson:94a,kaszkiel+Zobel:2001,Liu+Croft:02a,bendersky+Kurland:10}. The most commonly used passage-based document retrieval
methods rank a document by the highest query similarity exhibited by
any of its passages
\citep{callan:1994,Wilkinson:94a,kaszkiel+Zobel:2001,Liu+Croft:02a,bendersky+Kurland:10}
and by integrating this similarity with the document-query similarity
\citep{callan:1994,Wilkinson:94a,bendersky+Kurland:10}.

The passage-query (surface level) similarity is one out of many possible estimates for passage relevance. Indeed, various
passage-relevance estimates were devised for the task of passage
retrieval, a.k.a focused retrieval; e.g.,
\cite{Salton+al:93a,jiang+Zhai:2004,Murdock+Croft:2005,Murdock:2006,Metzler+Kanungo:2008,buffoni+al:2010,fernandez+al:2011,Fernandez+Losada:12a,carmel+al:2013,Keikha+al:14a,Chen+al:15a,yang+al:2016,Yulianti+al:16a,chen+al:2017}. That
is, passages are ranked in response to a query using
passage-relevance estimates. The merits of integrating the estimates
using learning-to-rank (LTR) approaches were also demonstrated \citep{Metzler+Kanungo:2008,buffoni+al:2010,Chen+al:15a,yang+al:2016,Yulianti+al:16a,chen+al:2017}.

Motivated by the (recent) progress in devising effective passage
retrieval methods, specifically, using LTR methods, and the fact that the main passage-based information used by most passage-based document retrieval methods is confined to passage-query similarities, we address the
following challenge: devising LTR methods for document retrieval that
utilize an effective query-based passage ranking. Some of the methods
we present are not based on any assumptions regarding the passage retrieval approach used to
rank passages. Others are based on the premise that passages were
ranked in response to the query using an LTR method that utilizes passage-based features. A
case in point, the most effective LTR-based document retrieval method
that we present uses both document-based and passage-based features;
the latter are those of the document's passage which is the most
highly ranked by an LTR method used to rank passages.

Each of the methods we present can be viewed as a conceptual analog,
or generalization, of previously proposed approaches for either (i)
passage-based document retrieval, where these approaches do not
utilize learning-to-rank or feature-based representations, or (ii)
cluster-based document retrieval.

In addition to presenting novel passage-based document retrieval
methods, we also propose new features for learning-to-rank
passages. These features are query-independent passage relevance
priors adapted from work on document retrieval over the Web
\citep{bendersky+al:2011}.

{
	Extensive empirical evaluation shows that our
	passage-based document retrieval approaches significantly outperform
	strong baselines. Further analysis demonstrates the importance of (i)
	utilizing an effective passage ranking, and (ii) using information
	induced from the document's passage that is the most highly ranked. In
	addition, we demonstrate the merits of using the query-independent passage
	features we propose for the task of passage retrieval. Specifically,
	integrating these features with previously proposed ones in a
	learning-to-rank approach results in passage retrieval performance
	that transcends the state-of-the-art.}

{Our contributions can be summarized as follows:
	\begin{itemize}
		\item A study of different methods that utilize passage-based features in a learning-to-rank approach for ranking documents.
		\item The utilization of an effective passage ranking for inducing document ranking, or in other words, addressing the question of how passage ranking can be transformed to document ranking.
		\item Some of our methods conceptually generalize previously
		proposed passage-based document retrieval methods which do not use learning-to-rank or feature-based representation.
		\item Some of our methods are conceptual reminiscent of cluster-based document retrieval approaches. This is the first work, to the best of our knowledge, to make the connection between passage-based document retrieval and cluster-based document retrieval.
		\item Attaining state-of-the-art retrieval performance across different collections and different feature sets.
		\item Demonstrating the effectiveness for passage retrieval of using passage-relevance priors adopted from work on document-relevance priors in Web retrieval.
	\end{itemize}
}

\section{Related Work}
The line of work most related to ours is on passage-based document
retrieval
\citep{Hearst+Plaunt:93a,callan:1994,Mittendorf+Schauble:1994,Wilkinson:94a,Kaszkiel+Zobel:97a,Denoyer+al:01a,kaszkiel+Zobel:2001,Liu+Croft:02a,Bendersky+Kurland:08a,Na+al:08a,Wang+al:08a,Wan+al:08a,bendersky+Kurland:10,Krikon+al:10a,Lang+al:10a}. 
As already noted, the most commonly used passage-based document retrieval methods are ranking a document by the
maximum query-similarity of its passages
\citep{callan:1994,Wilkinson:94a,Kaszkiel+Zobel:97a,kaszkiel+Zobel:2001,Liu+Croft:02a,Na+al:08a,bendersky+Kurland:10}
and by interpolating this similarity with the document-query
similarity
\citep{callan:1994,Wilkinson:94a,Na+al:08a,bendersky+Kurland:10}. 
We show that our best-performing
methods substantially outperform a highly effective method that integrates document-query and passage-query similarities \citep{bendersky+Kurland:10}.

In \cite{Wang+al:08a}, features based
on passage-query similarities were used to learn a document ranker
\citep{Wang+al:08a}. The induced ranking was fused with a
query-similarity-based document ranking. Our \fpd method, described in
Section \ref{sec:RetF}, generalizes this approach by using many
more passage features, integrating the resultant passage-based
document ranking with that produced by learning to rank documents, and
applying state-of-the-art learning-to-rank approaches. While \fpd is highly effective, it is outperformed by our best
performing method.

Recently, Yulianti et al. \cite{Yulianti+al:18a}
presented a method that selects (or generates) a passage from a
document in response to a query using information induced from a
community question answering system. Then, features of the passage
(not necessarily those used for selecting the passage) along with
document features are used to represent the document. This approach
is reminiscent of our \MethodThreeabb method which uses passage
features and document features to represent a document. There are,
however, major differences between the two. Our method is not based
on an external resource. Furthermore, we utilize passage ranking
that is induced using a learning-to-rank approach with passage
features while in \cite{Yulianti+al:18a} this is not the case. In
addition, the passage features used in our method are the same as
those used for ranking passages which is not the case in
Yulianti et al. \cite{Yulianti+al:18a}. We demonstrate the merits of using the
passage features that are used for (effective) passage ranking to
represent a document. We also show the merits of using passage-relevance prior estimates adopted from work on Web retrieval to rank passages. Some of these estimates were used by Yulianti+al et al. \cite{Yulianti+al:18a} to rank documents but not passages.

{Recently, a neural-network approach was presented for passage-based
	document retrieval \citep{Fan+al:18a}. Passage-query relevance signals
	(scores) are estimated using neural-network matching models and then
	aggregated to yield a document score. A difference with
	several of our models, in addition to using neural networks rather than
	a feature-based approach, is that ranking induced over passages from different documents is not utilized. A feature-based learning-to-rank baseline used in this work \citep{Fan+al:18a} represents a document using its features and the average, maximum and minimum values of query-similarities of its constituent passages. Therefore, this baseline is conceptually reminiscent of our \JPDm method which uses various aggregates of the feature values of document's passages together with the document features to represent documents. We show that there are passage-based features much more effective than passage-query similarities for estimating passage relevance, and accordingly, use aggregates of these features' values to represent documents.}

Some passage-based document retrieval methods use query
expansion \citep{Liu+Croft:02a,Lang+al:10a} or 
inter-passage similarities
\citep{Wan+al:08a,Wang+al:08a,Krikon+al:10a}. Integrating query
expansion and information induced from inter-passage similarities in
our approaches is an interesting future venue. 

Passage-based document retrieval approaches utilize term proximity information by the virtue of using passages. There are many other approaches for utilizing term proximities \citep{Metzler+Croft:2005,Metzler+Croft:07a,Tao+Zhai:07a,Lv+Zhai:2009,Zhao+Yun:09a,Lang+al:10a,Lv+Zhai:10a,Miao+al:12a}. We show that our best performing method outperforms a state-of-the-art term proximity model: the sequential dependence model from the Markov Random Field framework \citep{Metzler+Croft:2005}.

The vast majority of previous work on passage-based document retrieval
has focused on using passages marked prior to retrieval
time. There are some methods that simultaneously mark passages and use
them for retrieval \citep{Mittendorf+Schauble:1994,Denoyer+al:01a,kaszkiel+Zobel:2001}. Hence, our methods are not committed to a specific approach of
passage markup.

To implement and evaluate our passage-based document retrieval
methods, we use a passage ranking method that is based on
learning-to-rank. Some of the features we use for passage retrieval are adopted from work on retrieving
sentences to create snippets \citep{Metzler+Kanungo:2008} and retrieving
sentences (and more generally passages) as answers to non-factoid
questions \citep{Keikha+al:14a,Chen+al:15a,yang+al:2016}. We show that passage
retrieval performance can be significantly improved if
we also use query-independent passage relevance priors adapted from work on
devising document relevance priors for Web retrieval \citep{bendersky+al:2011}. Query-independent sentence priors different than ours, mainly 
based on opinion/sentiment analysis, were used in past work on
sentence retrieval \citep{Fernandez+Losada:12a}. More generally, there is a big body of work on retrieving passages; e.g., \cite{Salton+al:93a,Mittendorf+Schauble:1994,jiang+Zhai:2004,carmel+al:2013,Keikha+al:14a,Keikha+al:14b,chen+al:2017}. Our focus is different: we devise methods that utilize passage retrieval to improve document retrieval. Yet, we empirically show that the passage retrieval method we use in our document retrieval methods outperforms state-of-the-art passage retrieval approaches. Still, as already noted, our document retrieval methods are not committed to a specific passage retrieval approach.

\setlength{\abovedisplayskip}{2pt}
\setlength{\belowdisplayskip}{2pt}
\setlength{\abovedisplayshortskip}{2pt}
\setlength{\belowdisplayshortskip}{2pt} 
\section{Retrieval Framework}
\label{sec:RetF}
Our goal is to rank documents in corpus $\coll$ with respect to query
$\query$. We devise document retrieval methods that utilize information induced from document passages. We
assume that passages were marked in documents using some approach; $\passage \in \doc$ indicates that passage $\passage$ is part of document $\doc$. The retrieval methods we present are not dependent on the type of passages used. If $S$ is a document set, $G(S)$ denotes the ranked list of all passages of documents in $S$, where ranking was performed using some passage retrieval method.

Let $\topRetGroup$ be an initially retrieved document list produced in
response to $\query$ by using some retrieval method; e.g., in the
experiments reported in Section \ref{subsec:expSetup} we use standard
language-model-based retrieval. Then, a learning-to-rank (LTR)
method \citep{liu:2009} is used to re-rank $\topRetGroup$; the
resultant ranked list is denoted $\topRetLTRList$.  The only
assumption we make about the LTR method is that it uses a
feature-based vector representation, $\fVector{\doc}{\query}$, for
every pair of a document $\doc$ and the query $\query$.

We devise document ranking methods that re-rank $\topRetLTRList$
using information induced from the ranked list $\initPsgGroup$ of all passages
in documents in $\topRetLTRList$.\footnote{Note that these passages are also the passages of documents in $\topRetGroup$ since $\topRetLTRList$ is a re-rank of $\topRetGroup$.}
Some of the approaches we present do not depend on the passage ranking method used to produce $\initPsgGroup$. Others are based on the
assumption that the ranking is induced using an LTR approach applied to
passages; a pair of passage $\passage$ and query $\query$ is
represented using the feature vector
$\fVector{\passage}{\query}$. The basic premise is that effective passage ranking can be utilized to improve document ranking.

\subsection{Passage-Based Document Ranking}
We now present five passage-based document retrieval
approaches that can be used to re-rank $\topRetLTRList$. They mainly differ
by the way they utilize information about the ranking of passages in
$\initPsgGroup$. {These methods are either inspired by, or bear important connections to, existing passage-based and cluster ranking approaches.}
\subsubsection{A fusion-based approach}
\label{sec:fusion}
The first method we consider is conceptually reminiscent of a commonly used
passage-based document retrieval approach. The approach linearly interpolates
the document-query similarity score with the highest query
similarity score of a passage in the
document \citep{callan:1994,Wilkinson:94a,bendersky+Kurland:10}.

Here, instead of relying on query similarities, we use the ranking of
documents in $\topRetLTRList$ and that of the passages in
$\topPsgLTRList$ to induce document and passage retrieval scores,
respectively. Specifically, we apply the rank-to-score transformation
used in the highly effective reciprocal rank fusion
method \citep{cormack2009reciprocal}. That is, the score assigned to
item $\arbItem$, passage or document, with respect to the list $\arbList$ it is
in, $\topPsgLTRList$ or $\topRetLTRList$, is:
$$\genScoreList{\arbItem}{\arbList} \definedas \frac{1}{\rrfParam+\rankFunc{\arbItem}{\arbList}};$$
$\rankFunc{\arbItem}{\arbList}$ is $\arbItem$'s rank in $\arbList$;
the top item is at rank $1$; $\rrfParam$ is a free
parameter.

The final retrieval score of document $\doc$ ($\in \topRetLTRList$) is:
\begin{equation}
\label{eq:fuse}
Score(\doc;\query) \definedas \alpha \genScoreList{\doc}{\topRetLTRList} +
(1-\alpha) \max_{\passage \in \doc} \genScoreList{\passage}{\topPsgLTRList};
\end{equation}
$\alpha$ is a free parameter. Thus, $\doc$ is ranked high if
it was originally ranked high in $\topRetLTRList$ and at least one of its passages was ranked high in $\topPsgLTRList$. 

{The method just presented essentially applies the reciprocal rank fusion approach to fuse two rankings of the documents in $\topRetLTRList$ and is therefore denoted \firstmention{\RRFshort}. The first is the LTR-based ranking of $\topRetLTRList$. That is, documents are ranked using a ranking function learned based on document-only features. The second ranking is based on the
	highest rank in $\topPsgLTRList$ of a document's passage. In other words, the retrieval score of a document with respect to this ranking is based on the reciprocal rank of its passage that is the highest ranked. Note that the method is agnostic to the retrieval methods that were used to produce $\topRetLTRList$ and $\topPsgLTRList$; e.g., these need not even be LTR methods. All the method relies on is the ranking of documents and the ranking of passages of these documents.}

\subsubsection{Utilizing various passage-ranking statistics}
The \RRFshort method utilizes only the highest ranked passage of a
document to assign its final retrieval score in Equation \ref{eq:fuse}. The next
method, \firstmention{\MethodTwoTitle} (``statistics about multiple passages per
document''), ranks a document by utilizing various {\em statistics} regarding the ranking of the document's passages in $\topPsgLTRList$.

The feature vector used to represent a query-document pair is: 
$$\fVectorParam{\doc}{\query}{\MethodTwoTitle} \definedas \fVector{\doc}{\query} \oplus \fVectorTwo{\passage \in \doc}{\query}.$$
$\fVectorParam{\doc}{\query}{\MethodTwoTitle}$ is the concatenation of $\fVector{\doc}{\query}$: the original feature vector used to learn and apply the ranking function that served to induce $\topRetLTRList$ and $\fVectorTwo{\passage\in\doc}{\query}$: a vector composed of passage-based estimates. The estimates are the (i) maximum (max), (ii) minimum (min), (iii) average (avg), and
(iv) standard deviation (std) of $\genScoreList{\passage}{\topPsgLTRList}$ for
$\passage \in \doc$; (v) the fraction of passages in $\doc$ that are
among the $50$ (top50) and (vi) $100$ (top100) highest ranked passages in $\topPsgLTRList$;
and, (vii) the number of passages in $\doc$ (numPsg).

The rationale behind the \MethodTwoTitle method is to augment the
original document-query representation with ``statistics'' about the potential relevance
of its passages. The premise is that the relative ranking of passages in
$\topPsgLTRList$ can attest to their relevance to some extent. While \MethodTwoTitle is based on the fact that
$\topRetLTRList$ was indeed produced using an LTR approach, it is not
committed to a specific passage ranking method used to produce $\topPsgLTRList$.

{We note an interesting conceptual connection between \MethodTwoTitle and a cluster-based document retrieval method \citep{Kurland+Domshlak:08a}. The method ranks clusters of similar documents using measures that quantify the ranking of their constituent documents in a document ranking. In \MethodTwoTitle, we rank a document using measures that quantify the ranking of its constituent passages.}

\subsubsection{Joint passage-document representation using a single passage}
The next method, \firstmention{\MethodThreeabb} (``joint
passage document with a single passage''), similarly to the \RRFshort method,
uses $\doc$'s passage $\passage_{max}$ that is the
highest ranked in $\topPsgLTRList$. However, \MethodThreeabb does not rely on
$\passage_{max}$'s absolute rank in $\topPsgLTRList$, but only on the fact that it is the highest ranked among $\doc$'s passages. \MethodThreeabb is based on the premise that both $\topRetLTRList$ and $\topPsgLTRList$ were produced using LTR methods with feature vectors $\fVector{\doc}{\query}$ and $\fVector{\passage_{max}}{\query}$, respectively.
These two feature vectors are concatenated, and the resultant feature vector
$$\fVectorParam{\doc}{\query}{\MethodThreeabb} \definedas \fVector{\doc}{\query} \oplus \fVector{\passage_{max}}{\query},$$
is used for learning a ranker.

An important principle underlying \MethodThreeabb is to avoid {\em metric
	divergence} \citep {Metzler+Croft:2005}. That is, the features used to estimate the relevance of the document's passage that is presumably the most relevant --- according to $\topPsgLTRList$'s ranking --- are used directly, along with document-based features, to learn a document ranking function.

{\MethodThreeabb could be viewed as a conceptual generalization of the approach of smoothing a document language model with that induced from its passage which is the most similar to the query \citep{Bendersky+Kurland:08a}. That is, both approaches augment the document representation with information about its passage which is either the most query similar \citep{Bendersky+Kurland:08a} or the most highly ranked using a learning-to-rank approach (\MethodThreeabb). The difference is unsupervised method \citep{Bendersky+Kurland:08a} vs. a supervised method (\MethodThreeabb), and in accordance, representations (language models vs. feature vectors) and their integration (linear interpolation vs. concatenation).}

\subsubsection{Joint passage-document representation using multiple passages}
The \MethodThreeabb method uses information induced from a single
passage of $\doc$ to augment the document-query
feature-vector representation. We next consider an alternative, ``joint passage
document with multiple passages'' --- \firstmention{\JPDm} in short. The
document-query representation in \JPDm utilizes information
induced, potentially, from multiple passages.  Specifically, we define
a feature vector, $agg_{\passage \in \doc}(\fVector{\passage}{\query})$, based on the
same passage features used to represent passages in the LTR method that produced $\topPsgLTRList$. Each feature value in $agg_{\passage \in \doc}(\fVector{\passage}{\query})$ is the
aggregate of the corresponding feature values of all $\doc$'s passages. The feature
vector is then concatenated with the original document-query feature
vector
$$\fVectorParam{\doc}{\query}{\JPDm} \definedas \fVector{\doc}{\query} \oplus
agg_{\passage \in \doc}(\fVector{\passage}{\query});$$
$\fVectorParam{\doc}{\query}{\JPDm}$ is used for learning
a document ranking function.
The resultant methods are
termed \firstmention{\JPDAvg}, \firstmention{\JPDMax}
and \firstmention{\JPDMin} when using the average, maximum and minimum
aggregate functions, respectively. We note that \JPDm is the only approach we consider which does not use the ranking of passages in $\topPsgLTRList$.

It is important to highlight an additional difference between the \JPDm and \MethodTwoTitle methods, as both augment the document-query feature vector for learning a document ranking function with information induced from multiple passages in the document. While \MethodTwoTitle uses statistics mainly about the ranking of the document's passages, \JPDm utilizes passage-level features which were used to learn a passage ranker.

{There is an interesting conceptual connection between \JPDm and the ClustMRF method that ranks clusters of similar documents by the presumed percentage of relevant documents they contain \citep{Raiber+Kurland:13a}. In ClustMRF, clusters are represented using aggregates of feature values of their constituent documents  --- e.g., aggregates of document-query similarity scores, document-relevance prior estimates and more. \JPDm represents documents using aggregates of feature values of their constituent passages.}

{Finally, we note the important difference between \MethodThreeabb and \JPDm. In \MethodThreeabb, the passage-based features that are added to the
	document features represent a single passage; this is the document's
	most highly ranked passage. In contrast, in \JPDm, the passage-based
	features used to augment the document features do not represent a
	single passage: these are aggregates, over the document's passages, of
	feature values used in the passages' feature-vector representations. For example, in \JPDAvg, a single passage-based feature value would be the average feature value --- where average is computed over the document's passages --- for some feature in the feature-vector representation of the documents' passages. }

\subsubsection{Two-stage retrieval}
To further study the merits of simultaneously using document and
passage features to learn a document ranking function as in
the \MethodThreeabb and \JPDm methods presented above, we next explore
the \firstmention{\fpd} method (``first passage then document'').

A {\em document} ranking function is learned by representing the document-query pair with $\fVector{\passage_{max}}{\query}$ ---  the feature vector for the
document's 
passage $\passage_{max}$ that is the 
most highly ranked in $\topPsgLTRList$. That is, the learned document ranker utilizes only passage-based features. The ranker is then used to re-rank $\topRetLTRList$. The resultant ranking is fused
with $\topRetLTRList$'s original ranking using the reciprocal
rank approach as in \RRFshort. See Section \ref{sec:fusion} for further details\footnote{Experiments --- actual numbers are omitted as they convey no additional insight --- showed that simply
	using the passage-based document ranking without the additional fusion
	often yields performance (substantially) inferior to that of \fpd.}.

It is important to contrast the \fpd
and \RRFshort methods. Both fuse the original ranking of
$\topRetLTRList$ with a ranking based on utilizing passage-based information. The
difference is the type of passage-based information used.
While \RRFshort utilizes the rank in $\topPsgLTRList$ of the
document's most highly ranked passage to directly induce
document ranking, \fpd utilizes the passage-query feature vector of
this passage to learn and apply a document ranker.

We further note that \fpd depends on the fact that $\topPsgLTRList$
was induced using an LTR approach. In contrast, \fpd is not committed
to a specific retrieval method used to induce $\topRetLTRList$.

\section{Experimental Setting}
\label{subsec:expSetup}
The datasets used for experiments are specified in Table \ref{tab:datasets}.
\begin{table*}[t]
	\tabcolsep=0.05cm
	\caption{\label{tab:datasets} Datasets used for experiments.}
	\small
	\hspace*{-.1in}
	\begin{tabular}{@{}lcccc@{}}
		\toprule
		Corpus &   Data & \# of docs & Avg doc. length & Queries \\ 
		\midrule
		ROBUST & Disks 4\&5-CR & 528,155 & 479 & 301-450, 601-700\\ 
		WT10G & WT10g & 1,692,096 & 607 & 451-550 \\ 
		GOV2 & GOV2  & 25,205,179 & 930 & 701-850 \\ 
		ClueWeb & ClueWeb09 (Category B) & 50,220,423 & 807 & 1-200 \\ \midrule
		\inex & 2009\&2010 & 2,666,190 & 552 & 2009001-2009115, 2010001-2010107\\ 
		\aquaint & \aquaint & 1,033,461 & 436 & N1-N100 \\ 
		\bottomrule
	\end{tabular}
\end{table*}
ROBUST, WT10G, GOV2 and ClueWeb are TREC datasets.
ROBUST mostly contains newswire documents. WT10G is a small Web corpus. GOV2 is a crawl of the .gov domain. ClueWeb is a large-scale (noisy) Web collection. For {ClueWeb} we removed from the initial document rankings, described
below, documents with a Waterloo's spam classifier score below $50$ \citep{cormack+al:2011}.

The TREC datasets do not have passage-level relevance judgments that are needed for learning a passage-ranking method.
Thus, to learn a passage ranker we used the \inex dataset. The learned ranker was utilized by our passage-based document retrieval methods over all datasets.
The \inex dataset was used for the focused (passage) retrieval tracks in 2009 and 2010
\citep{geva+al:2010,arvola+al:2011}. It includes relevance judgments
for virtually every character in a relevant document; that is,
annotators marked the pieces of relevant text in relevant
documents. The dataset contains English Wikipedia documents from which we removed all XML tags; i.e., we treated the documents as plaintext. We use this dataset not only for learning a passage ranker, but also for evaluating the effectiveness of the learned ranker, as well as evaluating the effectiveness of our passage-based document retrieval methods in addition to the evaluation performed over the TREC datasets.

The passage features we propose are also used for learning and evaluating a passage ranker over the \aquaint{} collection which was
used for the novelty tracks in TREC 2003 and 2004
\citep{Soboroff+Harman:2003, Soboroff:2004}. In these tracks, relevant
documents have sentence-level relevance judgments. 
To perform sentence (passage) retrieval using the queries in both tracks, we follow the experimental setting in the 2003 track and rank the sentences in the set of relevant documents that were provided to participants.

Titles of topics served for queries. (Queries with no relevant
documents in the qrels were removed.)  The
Indri toolkit was used for all experiments\footnote{\url{www.lemurproject.org}}. We applied Krovetz stemming to
queries, documents (and their passages) and removed stopwords on the INQUERY list only
from queries. We used non-overlapping fixed-length windows
of $300$ terms for passages in our document retrieval methods. Such passages were shown to be effective for
passage-based document retrieval \citep{kaszkiel+Zobel:2001}. In Section \ref{sec:results} we study the effect of passage length on passage retrieval performance.

{Our main experiments are 
	conducted with} two learning-to-rank (LTR) methods for ranking documents and
passages: \LambdaMART \citep{Burges:2008} (LMart in
short)\footnote{\url{https://code.google.com/p/jforests/}.} or a
linear \SVM
\citep{joachims:2006}\footnote{\url{https://www.cs.cornell.edu/people/tj/svm_light/svm_rank.html}.}
{(SVM in short)}. \LambdaMART was trained for NDCG@10. {In Section \ref{sec:moreLTR} we present experimental results for two additional learning-to-rank methods.}

We measure the similarity between texts $x$ and $y$ (e.g., a query, a document or a passage) using the minus cross entropy between the unigram language models induced from them:
\begin{equation}
\label{eq:simFn}
\simFn{x}{y} \definedas \exp (-\ce{\theta_x^{MLE}}{\theta_y^{Dir}});
\end{equation}
$\theta_x^{MLE}$ is the unsmoothed maximum likelihood estimate induced from $x$ and $\theta_y^{Dir}$ is a Dirichlet smoothed language model induced from $y$ \citep{Zhai+Lafferty:2001}.

The two-tailed paired t-test with a $95\%$ confidence
level was used to determine statistically significant retrieval
performance differences. We applied Bonferroni correction for multiple hypothesis testing; i.e., when comparing a method with multiple baselines.

\subsection{Document Retrieval}
\label{sec:docRetEval}
We use a standard (unigram) language model approach 
(\firstmention{LM}) to retrieve an initial document list
$\topRetGroup$ of $1000$ documents for $\query$: document $\doc$ is scored by $\simFn{\query}{\doc}$.  
We then (re-)rank $\topRetGroup$ using an LTR method to obtain
$\topRetLTRList$; \firstmention{init-LTR} denotes this ranking.  Since
some of the datasets used for evaluation do not have hyperlink and
hypertext information, we only use highly effective content-based features. Specifically, the first three features
in the document-query feature vector $\fVector{\doc}{\query}$ are those of the sequential dependence
model (\SDM) from the Markov Random Field (MRF) framework
\citep{Metzler+Croft:2005}: unigrams, ordered bigrams and
unordered bigrams (biterms). \SDM is a state-of-the-art term-proximity model. The next three features are the most effective {\em document
	relevance priors} reported in \citep{bendersky+al:2011}:
(i) \firstmention{\swOne} and (ii) \firstmention{\swTwo} are the fraction
of terms in $\doc$ that are stopwords on the \inquery list, and the
fraction of stopwords on the \inquery list that appear in $\doc$
respectively, and (iii) the entropy, \firstmention{\entropy}, of the term
distribution in $\doc$.
High presence of
stopwords, and high entropy, presumably attests to rich use of language and therefore to content breadth \citep{bendersky+al:2011}.
{In Section \ref{sec:letor} we also present experimental results when using the \Mslr\footnote{\url{www.research.microsoft.com/en-us/projects/mslr}} features used in the LETOR datasets.}

The set of all passages in documents in
$\topRetLTRList$ is ranked to yield
$\topPsgLTRList$. The same LTR method used to produce $\topRetLTRList$ is used to produce $\topPsgLTRList$ with the passage-based features 
described  in Section \ref{sec:pRetF}. Then, $\topRetLTRList$ is re-ranked using the
document retrieval methods from Section \ref{sec:RetF} that utilize $\topPsgLTRList$. We use MAP and p@10 to evaluate document retrieval performance.

\myparagraph{Baselines} Recall that $\topRetLTRList$ was attained by
re-ranking $\topRetGroup$ using an LTR approach; i.e., the set of
documents in these two lists is the same. All the baselines we
describe and our passage-based document retrieval methods from Section \ref{sec:RetF} are used to rank this document set.

The initial language-model-based ranking of $\topRetGroup$, denoted \firstmention{LM}, is the first baseline.
The second is the initial LTR-based ranking of $\topRetLTRList$, init-LTR.  {MRF}'s \firstmention{\SDM} with its three features
\citep{Metzler+Croft:2005} also serves as a reference comparison. \SDM is a special case of the LTR method used to induce $\topRetLTRList$ where document
relevance priors are not used.  Another reference comparison is
\firstmention{\HLabb} \citep{bendersky+Kurland:10} where
document $\doc$ is scored with $\lambda
\simFn{\query}{\doc} +(1-\lambda) \max_{\passage \in \doc} \simFn
{\query}{\passage}$; the value of $\lambda$ is negatively
correlated with $\doc$'s length which serves as a document
homogeneity measure \citep{bendersky+Kurland:10}. \HLabb is an
effective representative of the approach of interpolating
document-query and passage-query similarity estimates
\citep{callan:1994,Wilkinson:94a,bendersky+Kurland:10}.

\subsection{Features for Learning to Rank Passages}
\label{sec:pRetF}
\label{sec:psgRet}
\label{sec:features}
All our passage-based document ranking approaches (except for JPDm) utilize a ranking of the documents' passages; i.e., the ranked list $\topPsgLTRList$. We now turn to describe the features used for learning a passage ranker. Some of these are novel to this study.
The features are estimates of passage $\passage$'s relevance to the query $\query$. Let $\docOf$ denote $\passage$'s ambient
document which we assume is part of a document set $\docList$
retrieved for $\query$. $\psgList$ denotes the set of passages of documents in
$\docList$. If $\docList$ is the set of documents in $\topRetLTRList$, the list we aim to re-rank, then $\psgList$ is $\initPsgGroup$.

The \firstmention{\passageCE} feature is the (normalized) passage-query similarity: $\frac{\simFn{\query}{\passage}}{\sum_{\passage' \in
		\psgList}\simFn{\query}{\passage'}}$. Since passages are relatively short, the ambient document can provide context in estimating query similarities (cf. \cite{Murdock:2006}): \firstmention{\documentSim} is 
$\frac{\simFn{\query}{\docOf}}{\sum_{\doc' \in \docList} \simFn{\query}{\doc'}}$. Additional
document-based 
features are the maximum, average,
and standard deviation of $\simFn{\query}{\passage'}$ for
$\passage' \in \docOf$: \firstmention{\pMax}, \firstmention{\pAvg} and
\firstmention{\std}, respectively. 
The longer $\passage$ is with respect to $\docOf$, the less reliance on
document-based query-similarity information is called for \citep{bendersky+Kurland:10}. Therefore, 
the ratio between $\passage$'s and $\docOf$'s lengths serves as a query independent feature: \firstmention{\lengthRatio}. 

Passages (if exist) that precede ($\passage_{pre}$) and follow
($\passage_{follow}$) $\passage$ in $\docOf$ provide focused context
for $\passage$ \citep{fernandez+al:2011}. Hence, we use \firstmention{\nighborLeft} and \firstmention{\neighborRight}: $\simFn{\query}{\passage_{pre}}$ and $\simFn{\query}{\passage_{follow}}$, respectively. If $\passage$ is the first or the last passage in the document, we use $\simFn{\query}{\passage}$ for \neighborLeft and \neighborRight, respectively.

The next features --- the use of which for passage retrieval is novel
to this study ---
are query-independent passage relevance priors. These are adopted from work on document relevance priors in Web search \citep{bendersky+al:2011}.
Specifically, we use the entropy (\entropy) and stopwords (\swOne, \swTwo) features described above, but now for passages rather than documents.

The passage independent
feature \firstmention{\queryTermCount} is the number of unique
terms in the query. This feature can potentially help to improve the
performance of non-linear rankers (cf., \cite{macdonald+al:2012}).

The next features are adopted from work on selecting sentences
for results' snippets \citep{Metzler+Kanungo:2008}. These were also used to
retrieve sentences (passages) for questions \citep{Chen+al:15a,yang+al:2016}.
\firstmention{\MKexactMatch} is true if $\query$ is a substring of $\passage$ and false otherwise. \firstmention{\MKtermOverlap} and \firstmention{\MKsynonymsOverlap} are the fraction of query terms and their synonyms (determined using Wordnet) in $\passage$. \firstmention{\MKgLength} is the number of terms in $\passage$ after removing stopwords, and \firstmention{\MKgLocation} is $\passage$'s position (in terms of passages) in $\docOf$ over the number of $\docOf$'s passages.

We also compare $\passage$ with $\query$ using the following three semantic-similarity
measures utilized for sentence-answer retrieval \citep{yang+al:2016}. (The first two were also used in \cite{Chen+al:15a}.)
The \firstmention{\ESA} similarity \citep{gabrilovich+Markovitch:2007}
is computed by using, separately, $\query$ and the $20$ terms in $\passage$ with
the highest TF.IDF values for query likelihood retrieval over the INEX Wikipedia
collection. The cosine measure is used to compare the lists of min-max normalized retrieval scores of the top-100 documents.

\firstmention{\SemWordEmbedding} is the cosine similarity
between the centroid of the Word2Vec 
vectors representing $\query$'s  terms and the centroid of the Word2Vec
vectors representing $\passage$'s terms. We used the 300 dimensional
newswire-based Word2Vec vectors from \url{https://code.google.com/p/word2vec/}.

\firstmention{\EntityLinking} is the Jaccard coefficient between the set-based entity representations of $\query$ and $\passage$. Wikipedia entities (i.e., titles) marked with a confidence level $ \ge 0.1$ by TagMe \citep{ferragina+Scaiella:2012} were used.

\subsubsection{Evaluating passage retrieval}
\label{sec:psgRetEval}
Most of our passage-based document ranking methods rely on the ranking of document passages. Hence, we also evaluate 
the effectiveness of the learned passage ranker using the \inex and \aquaint datasets --- this is a focused (passage) retrieval task. For \inex, the set $\initSet$, of all
passages of documents in the language-model-based initially retrieved document list $\topRetGroup$, is ranked; 
the top-$1500$ passages are evaluated using
\MAiP and \iP: precision at recall level $x \in\{.01,.1\}$ \citep{geva+al:2010,arvola+al:2011}. These evaluation measures were devised for the focused retrieval task where the percentage of relevant information in a passage is accounted for.
For \aquaint, following the novelty track in 2013 \citep{Soboroff+Harman:2003}, we set $\topRetGroup$ to be the provided set of relevant documents, and $\initSet$ is the set of all {\em sentences} in these documents which are ranked using our passage ranker. The top $1500$ ranked sentences are evaluated using \MAP and \PTen. (The tracks provided sentence-level binary relevance judgments.)

We use the following baselines
for passage ranking. The first method, \firstmention{\QuerySimFuse} (``query-similarity fusion'') \citep{callan:1994,carmel+al:2013}, 
scores $\passage$ 
by $(1-\lambda)\frac{\simFn{\query}{\passage}}{\sum_{\passage' \in \initSet}\simFn{\query}{\passage'}} +
\lambda \frac{\simFn{\query}{\docOf}}{\sum_{\doc' \in \topRetGroup}\simFn{\query}{\doc'}}$; $\lambda$ is a free
parameter.  
The two components of this interpolation are among the features used above for learning a passage ranker.

A tf.idf-based positional model was used for passage retrieval \citep{carmel+al:2013}. We use a language-model-based positional approach
\citep{Lv+Zhai:2009}, \firstmention{\PBCShort}, with a Gaussian
kernel, as other methods also utilize language models: $\passage$ is scored by $\lambda
\frac{\simFn{\query}{i_{max}(g)}}{\sum_{\passage'\in\initSet}\simFn{\query}{i_{max}(\passage')}} + 
\beta \frac{\simFn{\query}{\passage}}{\sum_{\passage' \in
		\initSet}\simFn{\query}{\passage'}}  +
(1-\lambda-\beta)\frac{\simFn{\query}{\docOf}}{\sum_{\doc' \in
		\topRetGroup}\simFn{\query}{\doc'}}$; $i_{max}(g)$ is the position
in $g$ whose Dirichlet induced language model yields the highest query
similarity among all positions $i$ in $g$; $\lambda$ and
$\beta$ are free parameters. Using \PBCShort as a feature in our passage ranking approach showed no merit.

We adapt the \firstmention{\owpc} method \citep{buffoni+al:2010}, originally used to rank structured XML elements, as an additional baseline.  For
compliance with our setting, all features except for those which rely
on XML structure are used in the two LTR methods used for all experiments. Most features rely on the query-similarity of
the passage and its ambient document; most of the features described above, which we use for learning a passage ranker, were not utilized.

The state-of-the-art LTR-based baseline, \firstmention{\MKS}, utilizes all the features proposed in \citep{yang+al:2016} for retrieving answer sentences to non-factoid questions. 
Our passage ranker utilizes some of these features.

The LTR-based approaches, \owpc, \MKS and our methods, are used to
re-rank the top $1500$ passages retrieved by \QuerySimFuse which is
considered an effective method. {Applying LTR methods
	on an initially retrieved list is common practice~\citep{liu:2009};
	specifically, the list size, for document retrieval, is often the
	same as that of the number of documents to be retrieved (e.g.,
	1000); hence, LTR methods often operate as re-ranking
	approaches. Similarly, the $1500$ threshold used here for passage
	retrieval corresponds to the standard passage list size used in the
	focused retrieval track of INEX \citep{geva+al:2010,arvola+al:2011}.}

\subsection{Additional Experimental Details}
As already noted, we use the INEX dataset to train a passage ranker with the features
described in Section \ref{sec:pRetF}. The ranker is also used for passage-based document retrieval over the TREC corpora which
lack focused (passage) relevance judgments. To learn a ranker, all passages of
documents in the initial language-model-based document list retrieved from INEX,
$\topRetGroup$, are ranked using the \QuerySimFuse method
described in Section \ref{sec:psgRetEval}; thus, $\topRetGroup$ serves for the set $\docList$ in Section \ref{sec:pRetF}. The top $1500$ passages serve for training. We explored a few binary/graded passage relevance-grade
definitions for learning a passage ranker. These use the fraction of relevant characters in a passage, denoted $\RFrac$.
A bucket-based approach which produces five relevance
grades resulted in effective performance of our passage ranker and the \owpc and \MKS baselines (see Section \ref{sec:psgRetEval} for details): $0$: $\RFrac <
.10$; $1$: $.10 \le \RFrac < .25$; $2$: $.25 \le \RFrac < .50$; $3$: $.50
\le \RFrac < .75$; $4$: $.75 \le \RFrac$.

To learn a passage ranking function for the sentence retrieval (ranking) task over AQUAINT, we use the sentences' binary relevance judgments as relevance grades. 

For the JPDs passage-based document retrieval approach, the DocQuerySim passage feature is not used, as it is the unigram feature of SDM that is used as a document-based feature. For the \JPDAvg and \JPDMax passage-based document retrieval methods, we do not use the passage-query similarity feature \passageCE (see Section \ref{sec:pRetF}) in $agg_{\passage \in \doc}(\fVector{\passage}{\query})$ since aggregating this feature value across the passages in the document amounts to the \pAvg and \pMax features, respectively, which are already used in $\fVector{\passage}{\query}$.

We used leave-one-out cross
validation over queries for training and testing; i.e., each query was used once for test wherein all other queries were used for training. 
For the LTR
methods we randomly split the train set to train ($80\%$) and validation
($20\%$);\footnote{The only exception was that the passage LTR method
	applied on TREC corpora was learned using all queries in the INEX
	dataset.} the latter was used to set the hyper parameters of the LTR
methods. For consistency, we use the same train set to set the free-parameter values of the non-LTR baselines (i.e., the validation set is not used for these methods).
MAP and \MAiP served as the optimization criteria for values of (hyper-) parameters in document and
passage retrieval, respectively. We min-max normalized the feature values used in the learning-to-rank methods on a per-query basis.

The Dirichlet smoothing parameter was set to $1000$
\citep{Zhai+Lafferty:2001} for the initial language-model-based document retrieval, and to
values in $\{500,1500,2500\}$ in all other cases.
The three parameters of \MRF's \SDM are set to values in
$\{0,0.1,\ldots,1\}$.
The 
value of $\lambda$ in 
\QuerySimFuse is in
$\{0.1,0.2,\ldots,0.9\}$. \SVM's regularization parameter is set to $\{0.0001,0.01,0.1\}$; all other hyper parameters of \SVM, and those of \LambdaMART, are set to 
default values of the implementations.

For \PLMShort, the value of the steepness parameter of the Gaussian kernel is in $\left\{50,100,\ldots,300\right\}$; $\lambda$ and $\beta$ were set to values in
{$\{0,0.2,\ldots,1\}$} \citep{Lv+Zhai:2009}. $\alpha$ (in the \RRFshort and \fpd methods from Section \ref{sec:RetF})  and $\rrfParam$ (in the \RRFshort, \MethodTwoTitle and \fpd methods from Section \ref{sec:RetF}) are in $\{0,0.1,\ldots,1\}$ and $\{0,30,60,90,100\}$, respectively.

\section{Experimental Results}
\label{sec:results}
In Section \ref{sec:psgDocRet} we analyze the performance of our passage-based document retrieval methods described in Section \ref{sec:RetF}.
As these methods rely on passage ranking, in Section \ref{sec:psgRetResults} we analyze the performance of our learning-to-rank-based passage retrieval method.
\subsection{Passge-Based Document Retrieval}
\label{sec:psgDocRet}

\begin{table*}[t]
	\tabcolsep=0.1cm
	\small
	\caption{\label{tab:Main_Result_JPDs_Baselines} Main result. Comparison between document retrieval baselines and \MethodThreeabb-LTR which is shown below to be our best performing method. '$\sigDoc$', '$\sigHome$', '$\sigMRF$' and '$\sigDocLTR$' mark statistically significant differences with \QL, \HLabb, SDM and \init-LTR respectively. Comparisons between LTR-based methods are performed between two methods utilizing the same LTR approach. Boldface: best result per column.}
	\center
	\begin{tabular}{@{}lcccccccccc@{}}
		\toprule
		\multicolumn{1}{l}{} & \multicolumn{2}{c}{ROBUST} 
		& \multicolumn{2}{c}{WT10G} & \multicolumn{2}{c}{GOV2} 
		& \multicolumn{2}{c}{ClueWeb} 
		& \multicolumn{2}{c}{INEX} \\ \cmidrule(l){2-3} \cmidrule(l){4-5} \cmidrule(l){6-7} \cmidrule(l){8-9} \cmidrule(l){10-11}
		& {\MAP} & {\PTen} & 
		{\MAP} & {\PTen} & 
		{\MAP} & {\PTen} & 
		{\MAP} & {\PTen} & 
		{\MAP} & {\PTen}\\
		\midrule
		
		
		\QL  &
		$.254$ & $.433$ &
		$.195$ & $.290$ & 
		$.292$ & $.534$ &
		$.187$ & $.339$ & 
		$.367$ & $.554$
		\\
		
		\HLabb  &
		$.254$ & $.424$ &
		$.209$ & $.292$ & 
		$.298$ & $.523$ &
		$.168$ & $.306$ & 
		$.368$ & $.538$
		\\
		
		\SDM  &
		$.261$ & $.440$ &
		$.202$ & $.293$ & 
		$.304$ & $.576$ &
		$.192$ & $.338$ & 
		$.385$ & $.568$ 
		\\
		
		\midrule
		init-\SVMabb  &
		$.261$ & $.439$ &
		$.213$ & $.334$ &
		$.336$ & $.643$ &
		$.222$ & $.406$ & 
		$.392$ & $.577$
		\\

		init-\LambdaMARTabb  &
		$.245$ & $.427$ &
		$.198$ & $.311$ &
		$.326$ & $.651$ &
		$.224$ & $.394$ & 
		$.378$ & $.584$
		\\ 
		
		\midrule
		
		\MethodThreeabb-\SVMabb  &
		
		$\mathbf{.290}^{\sigDoc\sigHome}_{\sigMRF\sigDocLTR}$ & $\mathbf{.480}^{\sigDoc\sigHome}_{\sigMRF\sigDocLTR}$ &
		
		$\mathbf{.235}^{\sigDoc\sigHome}_{\sigMRF\sigDocLTR}$ & $\mathbf{.381}^{\sigDoc\sigHome}_{\sigMRF\sigDocLTR}$ &
		
		$\mathbf{.350}^{\sigDoc\sigHome}_{\sigMRF\sigDocLTR}$ & $\mathbf{.656}^{\sigDoc\sigHome}_{\sigMRF}$ &
		
		$\mathbf{.246}^{\sigDoc\sigHome}_{\sigMRF\sigDocLTR}$ & $\mathbf{.452}^{\sigDoc\sigHome}_{\sigMRF\sigDocLTR}$ & 
		
		$\mathbf{.417}^{\sigDoc\sigHome}_{\sigMRF\sigDocLTR}$ & $.589^{\sigDoc\sigHome}$
		
		\\	
		
		\MethodThreeabb-\LambdaMARTabb  &
		
		$\mathbf{.290}^{\sigDoc\sigHome}_{\sigMRF\sigDocLTR}$ & $.471^{\sigDoc\sigHome}_{\sigMRF\sigDocLTR}$ &
		
		$.229^{\sigDoc}_{\sigDocLTR}$ & $.378^{\sigDoc\sigHome}_{\sigMRF\sigDocLTR}$ &
		
		$.345^{\sigDoc\sigHome}_{\sigMRF\sigDocLTR}$ & $.655^{\sigDoc\sigHome}_{\sigMRF}$ &
		
		$.234^{\sigDoc\sigHome}_{\sigMRF}$ & $.423^{\sigDoc\sigHome}_{\sigMRF}$ & 
		
		$.412^{\sigDoc\sigHome}_{\sigMRF\sigDocLTR}$ & $\mathbf{.593}^{\sigDoc\sigHome}$
		
		\\	  
		\bottomrule

	\end{tabular}
	
\end{table*}

\subsubsection{Main Result}
{
	Table \ref{tab:Main_Result_JPDs_Baselines} presents our main result.
	We see that in all relevant comparisons (5 datasets $\times$ 2
	evaluation measures), \MethodThreeabb, which is shown below to be our
	best performing approach, substantially outperforms all baselines: LM (unigram
	language-model-based retrieval), DocPsg (a representative
	passage-based document retrieval approach), SDM (a state-of-the-art
	term proximity method) and init-LTR (a learning-to-rank approach that
	utilizes document-query features). Most improvements are statistically
	significant. (We applied Bonferroni correction for multiple
	comparisons.) Refer back to Section \ref{sec:docRetEval} for more details about the baselines.}

{
	Recall that \MethodThreeabb learns a document ranker by utilizing
	the document-query features used to induce init-LTR and the
	passage-query features of the document's passage most highly ranked in
	response to the query. Its clear superiority with respect to the init-LTR methods attest to the merits of the
	way \MethodThreeabb leverages passage-based information.}

{ Given the performance superiority in most relevant comparisons of
	init-\SVMabb and init-\LambdaMARTabb to the other baselines, below
	we use them as reference comparisons. We note that their
	effectiveness attests to the effectiveness of the document features
	we use\footnote{The finding that init-\LambdaMARTabb underperforms
		init-\SVMabb can be attributed to the fact that \LambdaMARTabb is a
		non-linear ranker while \SVMabb is, and the number of queries used
		for training is not very large.} (See Section \ref{sec:docRetEval}
	for details regarding the features.)  }

{
	Since our methods
	utilize init-\SVMabb and init-\LambdaMARTabb (i.e., the initial list
	$\topRetLTRList$ or features used to induce it), and using each of
	the two entails a different experimental setting, we compare
	X-\SVMabb and X-\LambdaMARTabb methods separately.  }

\subsubsection{Comparing All Our Methods}
Table \ref{tab:PsgInfoDocRet} presents the performance comparison of
all our proposed passage-based document retrieval methods from Section
\ref{sec:RetF}. The init-LTR methods serve for reference comparison.

We see in Table \ref{tab:PsgInfoDocRet} that all the proposed methods outperform the init-LTR baselines  --- often
statistically significantly --- in the vast majority of relevant comparisons and are never outperformed in a
statistically significant manner by a baseline. 

\MethodThreeabb is the most effective approach among those
we proposed: its block in the table has the highest number of
boldfaced numbers, it outperforms any other approach in most relevance comparisons, and it is never statistically significantly
outperformed by other approaches while the reverse often holds. These findings attest to the merits of using the passage-query features of the document's passage most highly ranked together with the document-query features to learn a document ranker.

\begin{table*}[t]
	\tabcolsep=0.05cm
	
	\caption{\label{tab:PsgInfoDocRet}  Comparison of all our passage-based document retrieval methods. '$\sigDocLTR$' and '$\sigBP$' mark statistically significant differences with \init-LTR and \MethodThreeabb-LTR, respectively. Comparisons between LTR-based methods are performed between two methods utilizing the same LTR approach. Boldface: best result per column.}
	\center
	\small
	\begin{tabular}{@{}lcccccccccc@{}}
		\toprule
		\multicolumn{1}{l}{}  & \multicolumn{2}{c}{ROBUST} 
		& \multicolumn{2}{c}{WT10G} & \multicolumn{2}{c}{GOV2} 
		& \multicolumn{2}{c}{ClueWeb} & 
		\multicolumn{2}{c}{INEX} \\ \cmidrule(l){2-3} \cmidrule(l){4-5} \cmidrule(l){6-7} \cmidrule(l){8-9} \cmidrule(l){10-11}
		& {\MAP} & {\PTen} & 
		{\MAP} & {\PTen} & 
		{\MAP} & {\PTen} & 
		{\MAP} & {\PTen} & 
		{\MAP} & {\PTen}\\
		\midrule
		
		
		\init-\SVMabb &
		$.261$ & $.439$ &
		$.213$ & $.334$ &
		$.336$ & $.643$ &
		$.222$ & $.406$ & 
		$.392$ & $.577$
		\\

		\init-\LambdaMARTabb  &
		$.245$ & $.427$ &
		$.198$ & $.311$ &
		$.326$ & $.651$ &
		$.224$ & $.394$ & 
		$.378$ & $.584$
		\\ 
		\midrule
		
		\MethodThreeabb-\SVMabb  &
		
		$.290$ & $\mathbf{.480}$ &
		
		$.235$ & $\mathbf{.381}$ &
		
		$\mathbf{.350}$ & $.656$ &
		
		$\mathbf{.246}$ & $\mathbf{.452}$ & 
		
		$.417$ & $.589$
		
		\\	
		
		\MethodThreeabb-\LambdaMARTabb  &
		
		$.290$ & $.471$ &
		
		$.229$ & $.378$ &
		
		$.345$ & $.655$ &
		
		$.234$ & $.423$ & 
		
		$.412$ & $.593$
		
		\\	  \midrule
		
		\RRFshort-\SVMabb  &
		
		$.275^{\sigDocLTR\sigBP}$ & $.462^{\sigDocLTR\sigBP}$ &
		
		$.231^{\sigDocLTR}$ & $.376^{\sigDocLTR}$ &
		
		$.346^{\sigDocLTR}$ & $.639$ &
		
		$.234^{\sigDocLTR\sigBP}$ & $.425^{\sigDocLTR\sigBP}$ & 
		
		$.408^{\sigDocLTR\sigBP}$ & $.601^{\sigDocLTR}$
		
		\\
		
		\RRFshort-\LambdaMARTabb &
		
		$.281^{\sigDocLTR\sigBP}$ & $.462^{\sigDocLTR}$ &
		
		$.230^{\sigDocLTR}$ & $.367^{\sigDocLTR}$ &
		
		$.339^{\sigDocLTR\sigBP}$ & $.638$ &
		
		$.232^{\sigDocLTR}$ & $.427^{\sigDocLTR}$ & 
		
		$.410^{\sigDocLTR}$ & $.603$
		
		\\ 
		
		\midrule

		\MethodTwoTitle-\SVMabb  &
		
		$.271^{\sigDocLTR\sigBP}$ & $.455^{\sigDocLTR\sigBP}$ &
		
		$.223^{\sigDocLTR\sigBP}$ & $.363^{\sigDocLTR}$ &
		
		$.344^{\sigDocLTR\sigBP}$ & $.647$ &
		
		$.233^{\sigDocLTR\sigBP}$ & $.418^{\sigBP}$ & 
		
		$.401^{\sigDocLTR\sigBP}$ & $.598^{\sigDocLTR}$
		
		\\ 
		
		\MethodTwoTitle-\LambdaMARTabb  &
		
		$.280^{\sigDocLTR\sigBP}$ & $.460^{\sigDocLTR}$ &
		
		$\mathbf{.236^{\sigDocLTR}}$ & $.370^{\sigDocLTR}$ &
		
		$.341^{\sigDocLTR}$ & $.641$ &
		
		$.239^{\sigDocLTR}$ & $.433^{\sigDocLTR}$ & 
		
		$.412^{\sigDocLTR}$ & $.600$
		\\  
		
		\midrule

		\JPDAvg-\SVMabb   & 
		
		$.285^{\sigDocLTR\sigBP}$ & $.465^{\sigDocLTR\sigBP}$ &
		
		$.228^{\sigDocLTR}$ & $.363^{\sigDocLTR}$ &
		
		$.343^{\sigBP}$ & $.639$ &
		
		$.244^{\sigDocLTR}$ & $.434^{\sigDocLTR\sigBP}$ & 
		
		$.415^{\sigDocLTR}$ & $.598^{\sigDocLTR}$
		
		\\ 
		\JPDAvg-\LambdaMARTabb  &
		
		$.288^{\sigDocLTR}$ & $.471^{\sigDocLTR}$ &
		
		$.223^{\sigDocLTR}$ & $.355^{\sigDocLTR\sigBP}$ & 
		
		$.342^{\sigDocLTR}$ & $\mathbf{.663}$ &
		
		$.237^{\sigDocLTR}$ & $.422^{\sigDocLTR}$ & 
		
		$.417^{\sigDocLTR}$ & $.595$
		
		\\ \midrule
		
		\JPDMax-\SVMabb   &
		
		$\mathbf{.293}^{\sigDocLTR}$ & $.476^{\sigDocLTR}$ &
		
		$.235^{\sigDocLTR}$ & $.374^{\sigDocLTR}$ & 
		
		$\mathbf{.350}^{\sigDocLTR}$ & $.643$ &
		
		$.242^{\sigDocLTR}$ & $.429^{\sigBP}$ & 
		
		$\mathbf{.420^{\sigDocLTR}}$ & $.601^{\sigDocLTR}$
		
		\\
		\JPDMax-\LambdaMARTabb  &
		
		$.289^{\sigDocLTR}$ & $.468^{\sigDocLTR}$ & 
		
		$.228^{\sigDocLTR}$ & $.363^{\sigDocLTR}$ &
		
		$.349^{\sigDocLTR}$ & $.654$ &
		
		$.230$ & $.416$ & 
		
		$.416^{\sigDocLTR}$ & $.602$
		
		\\ \midrule
		
		\JPDMin-\SVMabb   & 
		
		$.270^{\sigDocLTR\sigBP}$ & $.451^{\sigBP}$ &
		
		$.233^{\sigDocLTR}$ & $.342^{\sigBP}$ &
		
		$.334^{\sigBP}$ & $.630^{\sigBP}$ &
		
		$.236^{\sigDocLTR}$ & $.430^{\sigDocLTR\sigBP}$ & 
		
		$.404^{\sigDocLTR\sigBP}$ & $.583$
		
		\\ 
		\JPDMin-\LambdaMARTabb  &
		
		$.271^{\sigDocLTR\sigBP}$ & $.454^{\sigDocLTR\sigBP}$ &
		
		$.220^{\sigDocLTR}$ & $.338^{\sigBP}$ &
		
		$.337^{\sigDocLTR\sigBP}$ & $.640$ &
		
		$.230$ & $.403^{\sigBP}$ & 
		
		$.394^{\sigDocLTR\sigBP}$ & $.578$
		
		\\ \midrule

		\fpd-\SVMabb  &
		
		$.288^{\sigDocLTR\sigBP}$ & $.474^{\sigDocLTR}$ &
		
		$.228^{\sigDocLTR\sigBP}$ & $.372^{\sigDocLTR}$ &
		
		$.348^{\sigDocLTR}$ & $.643$ &
		
		$.238^{\sigDocLTR\sigBP}$ & $.434^{\sigDocLTR\sigBP}$ & 
		
		$.411^{\sigDocLTR\sigBP}$ & $.588$
		
		\\	
		
		\fpd-\LambdaMARTabb  &
		
		$.291^{\sigDocLTR}$ & $.468^{\sigDocLTR}$ &
		
		$.228^{\sigDocLTR}$ & $.362^{\sigDocLTR}$ &
		
		$.349^{\sigDocLTR}$ & $.655$ &
		
		$.236^{\sigDocLTR}$ & $.423^{\sigDocLTR}$ & 
		
		$.414^{\sigDocLTR}$ & $\mathbf{.609^{\sigDocLTR}}$
		
		\\	
		\bottomrule

	\end{tabular}
	
\end{table*}

The JPDm-max approach is the second-best performing.
This finding is not entirely surprising: \MethodThreeabb, which is our best performing method, uses the features of the document's passage most highly ranked while JPDm-max uses per each passage-based feature the maximum value over the document's passages.
As could be expected, both JPDm-max and JPDm-avg outperform JPDm-min. That is, using the average or the maximum of a feature value across the document's passages yields better performance than using the minimal value.

Table \ref{tab:PsgInfoDocRet} also shows that \RRFshort
outperforms \MethodTwoTitle in most relevant comparisons when using SVM and the reverse holds when using LMart. {However, only the \MAP differences between \RRFshort-\SVMabb and \MethodTwoTitle-\SVMabb for ROBUST and INEX are statistically significant.}
We thus conclude
that the most important passage-rank-based information is the rank of a document's most highly ranked passage. (Recall that \MethodTwoTitle uses additional statistics about the ranking of passages of a document.) We attribute these findings to the fact that a document can be deemed relevant even if it contains only a single short relevant passage.

Another observation that we make based on Table \ref{tab:PsgInfoDocRet} is that \fpd and \MethodThreeabb  outperform \RRFshort in most relevant
comparisons; i.e., using the query-passage features of the passage
most highly ranked of a document is more effective than using its rank. Using these features together with document features
(\MethodThreeabb) is more effective than using them separately (\fpd)
to induce document ranking.

\subsubsection{Further Analysis of JPDs}
We saw above that JPDs is the most effective passage-based document retrieval approach among those we proposed. JPDs uses together the document-query features and the passage-query features of the document's most highly ranked passage so as to learn a document ranking function. In Table \ref{tab:JPD_Versions_AllCollections} we contrast the performance of JPDs with that of its variants that use the passage-query features of the document's second (\firstmention{JPDs-second}), third (\firstmention{JPDs-third}) and lowest (\firstmention{JPDs-lowest}) ranked passages in $\topPsgLTRList$.

Table \ref{tab:JPD_Versions_AllCollections} shows that the
original version, JPDs, outperforms in most relevant comparisons its
variants (JPDs-second, JPDs-third and JPDs-lowest). More generally, we see that for almost all datasets, the lower the document's passage, whose passage-query features are used, is ranked, the lower the retrieval performance of the JPDs approach that uses these features.\footnote{We note that the use of the lowest ranked passage did not result in substantial performance decrease due to the length of passages used here: $300$; that is, such passages can incorporate a descent amount of information from the entire document, especially in cases of relatively short documents.} These findings attest to the merits of using the features of the document's most highly ranked passage, and to the fact that the relative ranking of the document's passages with respect to the query can imply to the benefits of using information induced from them to rank the document.

\begin{table*}[t]
	\tabcolsep=0.1cm
	
	\caption{\label{tab:JPD_Versions_AllCollections} Comparing variants of JPDs. Boldface: the best result in a column for each LTR method (SVM or LMart). '$\sigBP$' marks statistically significant differences with \MethodThreeabb-LTR.}	
	\center
	\small
	\begin{tabular}{@{}lcccccccccc@{}}
		\toprule
		\multicolumn{1}{l}{} & \multicolumn{2}{c}{ROBUST} 
		& \multicolumn{2}{c}{WT10G} & \multicolumn{2}{c}{GOV2} 
		& \multicolumn{2}{c}{ClueWeb} & 
		\multicolumn{2}{c}{INEX}\\ \cmidrule(l){2-3} \cmidrule(l){4-5} \cmidrule(l){6-7} \cmidrule(l){8-9} \cmidrule(l){10-11}
		& \MAP & \PTen & 
		\MAP & \PTen & 
		\MAP & \PTen & 
		\MAP & \PTen & 
		\MAP & \PTen  \\
		\midrule
		
		\MethodThreeabb-\SVMabb   &
		
		$\mathbf{.290}$ & $\mathbf{.480}$ & 
		
		$.235$ & $\mathbf{.381}$ &
		
		$\mathbf{.350}$ & $\mathbf{.656}$ &
		
		$\mathbf{.246}$ & $\mathbf{.452}$ & 
		
		$\mathbf{.417}$ & $.589$
		
		\\
		\midrule
		
		\JPDSecond-\SVMabb   & 
		
		$.277^{\sigBP}$ & $.464^{\sigBP}$ &
		
		$\mathbf{.236}$ & $.363$ &
		
		$.341^{\sigBP}$ & $.646$ &
		
		$.238^{\sigBP}$ & $.430^{\sigBP}$ & 
		
		$.414$ & $.594$
		
		\\
		
		\JPDThird-\SVMabb  &
		
		$.273^{\sigBP}$ & $.455^{\sigBP}$ &
		
		$.232$ & $.363$ &
		
		$.338^{\sigBP}$ & $.633^{\sigBP}$ &
		
		$.238^{\sigBP}$ & $.434^{\sigBP}$ & 
		
		$.412$ & $\mathbf{.598}$
		
		\\

		\JPDLast-\SVMabb   & 
		
		$.271^{\sigBP}$ & $.452^{\sigBP}$ &
		
		$.231$ & $.347^{\sigBP}$ &
		
		$.335^{\sigBP}$ & $.629^{\sigBP}$ & 
		
		$.226^{\sigBP}$ & $.410^{\sigBP}$  & 
		
		$.402^{\sigBP}$ & $.577$
		
		\\
		
		\midrule
		
		\MethodThreeabb-\LambdaMARTabb  & 
		
		$\mathbf{.290}$ & $\mathbf{.471}$ &
		
		$\mathbf{.229}$ & $\mathbf{.378}$ &
		
		$\mathbf{.345}$ & $\mathbf{.655}$ &
		
		$.234$ & $.423$  & 
		
		$\mathbf{.412}$ & $\mathbf{.593}$
		
		\\	
		\midrule
		
		\JPDSecond-\LambdaMARTabb  &
		
		$.280^{\sigBP}$ & $.458$ &
		
		$.226$ & $.358$ &
		
		$.341$ & $\mathbf{.655}$ &
		
		$\mathbf{.240}$ & $\mathbf{.429}$  & 
		
		$.410$ & $.588$
		
		\\	
		
		\JPDThird-\LambdaMARTabb & 
		
		$.273^{\sigBP}$ & $.455^{\sigBP}$ &
		
		$.218$ & $.361$ & 
		
		$.341$ & $.650$ &
		
		$.235$ & $.422$ & 
		
		$.401^{\sigBP}$ & $.587$ 
		
		\\
		
		\JPDLast-\LambdaMARTabb  &
		
		$.270^{\sigBP}$ & $.448^{\sigBP}$ &
		
		$.219$ & $.336^{\sigBP}$ &
		
		$.337^{\sigBP}$ & $.649$ &
		
		$.232$ & $.411$ & 
		
		$.400^{\sigBP}$ & $.581$
		
		\\
		
		\bottomrule

	\end{tabular}
	
\end{table*}
\begin{table*}[t]
	\tabcolsep=0.1cm
	\caption{\label{tab:JPDsVsJPDsTopK} {Comparing JPDs with JPD-2 where the features of the document's two most highly ranked passages are used in addition to those of the document. Boldface: the best result in a column for each LTR method (SVM or LMart). '$\sigBP$' marks statistically significant differences with \MethodThreeabb-LTR.}}	
	\center
	\small
	\begin{tabular}{@{}lcccccccccc@{}}
		\toprule
		\multicolumn{1}{l}{} & \multicolumn{2}{c}{ROBUST} 
		& \multicolumn{2}{c}{WT10G} & \multicolumn{2}{c}{GOV2} 
		& \multicolumn{2}{c}{ClueWeb} & 
		\multicolumn{2}{c}{INEX}\\ \cmidrule(l){2-3} \cmidrule(l){4-5} \cmidrule(l){6-7} \cmidrule(l){8-9} \cmidrule(l){10-11}
		& \MAP & \PTen & 
		\MAP & \PTen & 
		\MAP & \PTen & 
		\MAP & \PTen & 
		\MAP & \PTen  \\
		\midrule
		
		\MethodThreeabb-\SVMabb   &
		
		$.290$ & $\mathbf{.480}$ & 
		
		$\mathbf{.235}$ & $\mathbf{.381}$ &
		
		$.350$ & $\mathbf{.656}$ &
		
		$.246$ & $\mathbf{.452}$ & 
		
		$.417$ & $.589$
		
		\\
		
		JPD-2-\SVMabb   & 
		
		$\mathbf{.291}$ & $.473$ &
		
		$\mathbf{.235}$ & $.373$ &
		
		$\mathbf{.351}$ & $.655$ &
		
		$\mathbf{.250}^{\sigBP}$ & $\mathbf{.452}$  & 
		
		$\mathbf{.421}$ & $\mathbf{.601}$
		
		\\
		
		\midrule
		
		\MethodThreeabb-\LambdaMARTabb  & 
		
		$.290$ & $.471$ &
		
		$.229$ & $\mathbf{.378}$ &
		
		$.345$ & $\mathbf{.655}$ &
		
		$.234$ & $.423$  & 
		
		$.412$ & $.593$
		\\

		JPD-2-\LambdaMARTabb  &
		
		$\mathbf{.291}$ & $\mathbf{.473}$ &
		
		$\mathbf{.236}$ & $.375$ &
		
		$\mathbf{.349}$ & $.649$ &
		
		$\mathbf{.235}$ & $\mathbf{.430}$  & 
		
		$\mathbf{.418}$ & $\mathbf{.605}$
		
		\\	
		\bottomrule 
	\end{tabular}
	
\end{table*}

\begin{table*}[t]
	\tabcolsep=0.1cm
	\caption{\label{tab:GinitGltrComp} The effect on document ranking effectiveness of the passage ranker: LTR-based (\psgLTR) vs. integrating the passage-query similarity with the query-similarity of the passage's ambient document (\psgLM). '\sigRanking' marks statistically significant differences
		between \psgLTR and \psgLM. Boldface: the best result for evaluation measure in a block.}  \center
	\small
	\begin{tabular}{@{}lcccccccccc@{}}
		\toprule
		\multicolumn{1}{l}{}  & \multicolumn{2}{c}{ROBUST} 
		& \multicolumn{2}{c}{WT10G} & \multicolumn{2}{c}{GOV2} 
		& \multicolumn{2}{c}{ClueWeb} & 
		\multicolumn{2}{c}{INEX} \\ \cmidrule(l){2-3} \cmidrule(l){4-5} \cmidrule(l){6-7} \cmidrule(l){8-9} \cmidrule(l){10-11}
		& {\MAP} & {\PTen} & 
		{\MAP} & {\PTen} & 
		{\MAP} & {\PTen} & 
		{\MAP} & {\PTen} & 
		{\MAP} & {\PTen}\\
		\midrule
		
		\RRFshort-\SVMabb \psgLTR  &
		
		$\mathbf{.275}^{\sigRanking}$ & $\mathbf{.462}^{\sigRanking}$ &
		
		$\mathbf{.231}^{\sigRanking}$ & $\mathbf{.376}^{\sigRanking}$ &
		
		$\mathbf{.346}^{\sigRanking}$ & $.639$ &
		
		$\mathbf{.234}^{\sigRanking}$ & $\mathbf{.425}^{\sigRanking}$ & 
		
		$\mathbf{.408}^{\sigRanking}$ & $\mathbf{.601}^{\sigRanking}$
		
		\\
		
		\RRFshort-\SVMabb \psgLM  &
		
		$.261$ & $.442$ &
		
		$.215$ & $.324$ &
		
		$.336$ & $\mathbf{.643}$ &
		
		$.223$ & $.406$ &
		
		$.390$ & $.574$ 
		
		\\  \midrule
		
		\RRFshort-\LambdaMARTabb \psgLTR  &
		
		$\mathbf{.281}^{\sigRanking}$ & $\mathbf{.462}^{\sigRanking}$ &
		
		$\mathbf{.230}^{\sigRanking}$ & $\mathbf{.367}^{\sigRanking}$ &
		
		$\mathbf{.339}^{\sigRanking}$ & $.638$ &
		
		$\mathbf{.232}^{\sigRanking}$ & $\mathbf{.427}^{\sigRanking}$ & 
		
		$\mathbf{.410}^{\sigRanking}$ & $\mathbf{.603}^{\sigRanking}$
		
		\\ 
		
		\RRFshort-\LambdaMARTabb \psgLM  &
		
		$.257$ & $.442$ &
		
		$.204$ & $.318$ &
		
		$.326$ & $\mathbf{.645}$ &
		
		$.224$ & $.396$ &
		
		$.382$ & $.581$ 
		
		\\ 
		
		\midrule

		\MethodTwoTitle-\SVMabb \psgLTR  &
		
		$\mathbf{.271}^{\sigRanking}$ & $\mathbf{.455}^{\sigRanking}$ &
		
		$\mathbf{.223}^{\sigRanking}$ & $\mathbf{.363}^{\sigRanking}$ &
		
		$\mathbf{.344}$ & $\mathbf{.647}$ &
		
		$\mathbf{.233}$ & $\mathbf{.418}$ & 
		
		$\mathbf{.401}^{\sigRanking}$ & $\mathbf{.598}^{\sigRanking}$
		
		\\
		
		\MethodTwoTitle-\SVMabb  \psgLM  &
		
		$.259$ & $.439$ &
		
		$.213$ & $.337$ &
		
		$.337$ & $.642$ &
		
		$.227$ & $.409$ & 
		
		$.386$ & $.564$ 
		
		\\
		\midrule
		
		\MethodTwoTitle-\LambdaMARTabb \psgLTR  &
		
		$\mathbf{.280}^{\sigRanking}$ & $\mathbf{.460}^{\sigRanking}$ &
		
		$\mathbf{.236}^{\sigRanking}$ & $\mathbf{.370}^{\sigRanking}$ &
		
		$\mathbf{.341}$ & $.641$ &
		
		$\mathbf{.239}^{\sigRanking}$ & $\mathbf{.433}^{\sigRanking}$ & 
		
		$\mathbf{.412}^{\sigRanking}$ & $\mathbf{.600}$
		
		\\
		
		\MethodTwoTitle-\LambdaMARTabb \psgLM &
		
		$.258$ & $.442$ &
		
		$.211$ & $.327$ &
		
		$.336$ & $\mathbf{.651}$ &
		
		$.223$ & $.407$ & 
		
		$.389$ & $.579$ 
		
		\\
		\midrule
		
		\MethodThreeabb-\SVMabb \psgLTR  &
		
		$\mathbf{.290}$ & $\mathbf{.480}$ &
		
		$\mathbf{.235}$ & $\mathbf{.381}$ &
		
		$\mathbf{.350}$ & $\mathbf{.656}$ &
		
		$\mathbf{.246}$ & $\mathbf{.452}$ & 
		
		$\mathbf{.417}$ & $.589$
		
		\\
		
		\MethodThreeabb-\SVMabb \psgLM  &
		
		$.288$ & $.474$ &
		
		$.233$ & $.373$ &
		
		$.347$ & $.647$ &
		
		$.245$ & $.441$  & 
		
		$.414$ & $\mathbf{.595}$ 
		
		\\
		\midrule

		\MethodThreeabb-\LambdaMARTabb \psgLTR  &
		
		$\mathbf{.290}$ & $.471$ &
		
		$.229$ & $\mathbf{.378}$ &
		
		$\mathbf{.345}$ & $\mathbf{.655}$ &
		
		$\mathbf{.234}$ & $\mathbf{.423}$ & 
		
		$\mathbf{.412}$ & $.593$
		
		\\
		
		\MethodThreeabb-\LambdaMARTabb \psgLM  &
		
		$.289$ & $\mathbf{.473}$ &
		
		$\mathbf{.230}$ & $.365$ &
		
		$.343$ & $.641$ &
		
		$.228$ & $.407$  & 
		
		$.410$ & $\mathbf{.597}$ 
		
		\\
		
		\midrule
		
		\fpd-\SVMabb \psgLTR  &
		
		$\mathbf{.288}$ & $.474$ &
		
		$.228$ & $\mathbf{.372}$ &
		
		$\mathbf{.348}^{\sigRanking}$ & $\mathbf{.643}$ &
		
		$\mathbf{.238}^{\sigRanking}$ & $\mathbf{.434}$ & 
		
		$\mathbf{.411}^{\sigRanking}$ & $.588$
		
		\\
		
		\fpd-\SVMabb \psgLM  &
		
		$.286$ & $\mathbf{.475}$ &
		
		$\mathbf{.230}$ & $.365$ &
		
		$.345$ & $.632$ &
		
		$.230$ & $.422$ & 
		
		$.405$ & $\mathbf{.591}$ 
		
		\\
		\midrule

		\fpd-\LambdaMARTabb \psgLTR  &
		
		$\mathbf{.291}$ & $\mathbf{.468}$ &
		
		$\mathbf{.228}$ & $\mathbf{.362}$ &
		
		$\mathbf{.349}^{\sigRanking}$ & $\mathbf{.655}^{\sigRanking}$ &
		
		$\mathbf{.236}$ & $\mathbf{.423}$ & 
		
		$\mathbf{.414}$ & $\mathbf{.609}$  
		
		\\	
		
		\fpd-\LambdaMARTabb \psgLM &
		
		$.287$ & $\mathbf{.468}$ &
		
		$.225$ & $.361$ &
		
		$.344$ & $.631$ &
		
		$.233$ & $.417$ & 
		
		$.411$ & $.605$  
		
		\\	
		
		\bottomrule 
	\end{tabular}
	
\end{table*}

\subsubsection{Utilizing Two Passages}
\label{sec:twoPsg}
Our JPDs method utilizes the features of the document's most highly
ranked passage in addition to the document's features. We now consider
a variant of JPDs, denoted \firstmention{JPD-2}, which uses in
addition the features of the document's passage which is the second
ranked\footnote{To avoid having the same features used for the two passages, the following
	features were removed from the feature vector of the second ranked
	passage: \documentSim, \pMax, \pAvg, \std and
	\queryTermCount.}. The feature vectors of the two passages are
concatenated with that of the document for learning a document
ranker. Table \ref{tab:JPDsVsJPDsTopK} presents the results.

We see in Table \ref{tab:JPDsVsJPDsTopK} that using the two passages (JPD-2-LTR) yields performance that is very similar in most relevant comparisons to that of using a single passage (JPDs-LTR). In only a single case, the performance difference is statistically significant.

\subsubsection{The Effect of the Passage Ranker}
Our passage-based document retrieval approaches (except for \JPDm)
utilize information induced from the ranking of passages in the
initially retrieved document list, $\topRetGroup$.  In Table
\ref{tab:GinitGltrComp} we compare the performance of the approaches when
using two different passage ranking methods. The first is the QSF
method described in Section \ref{sec:psgRetEval} which integrates the
passage-query similarity value with the query-similarity value of the
passage's ambient document. The second passage
ranking method, \firstmention{\psgLTR}, was used
insofar: SVM or LMart applied with our proposed passage-based features
from Section \ref{sec:pRetF}\footnote{We do not present the comparison
	for the \JPDm approach as it is independent of the passage
	ranking.}. In Section \ref{sec:psgRetResults} we show that the passage-ranking effectiveness of
\psgLTR is substantially better than that of \psgLM.

The message rising from Table \ref{tab:GinitGltrComp} is clear: our passage-based document retrieval methods post better performance when using the LTR-based passage ranker than when using the \psgLM method to rank passages. While most improvements are statistically significant, those for JPDs are not. This finding attests to the robustness of JPDs with respect to the passage ranker used.

\subsubsection{Feature Analysis for Document Retrieval}
We now present feature analysis for our best performing approach, \MethodThreeabb. We start by analyzing \MethodThreeabb-SVM which outperforms \MethodThreeabb-LMart (see Table \ref{tab:Main_Result_JPDs_Baselines}). 

First, we average, per dataset, the
weights assigned to features in \MethodThreeabb-SVM using the different training folds. (Recall
that we use leave-one-out cross validation.) Then, the features are ordered in descending order of these averages. Each feature is assigned a score which is the reciprocal of its rank position in the ordered list. Finally,
features are ordered by averaging their scores across datasets. The top
$10$ features\footnote{\MethodThreeabb-SVM uses 24 features and \MethodThreeabb-LMart uses 25 features --- the additional feature is the query length which is not useful for a linear ranker.} according to this analysis are (p and d indicate that the feature is of the passage or the document, respectively): SDM unigrams (d), \ESA (p), \EntityLinking (p), \entropy (d), \pAvg (p),
\pMax (p), SW2 (d), SDM biterms (d), \MKsynonymsOverlap (p), \SemWordEmbedding (p). Thus, both document-based and passage-based features are among the top-5 and top-10. This finding attests to the merits of using both types of features to learn a document ranking function.

We also performed ablation tests for \MethodThreeabb where we removed one feature at a time. Actual numbers are omitted as they convey no additional insight. We order the features in descending order of the number of cases where their removal resulted in statistically significant performance drop. A case is defined by a dataset and evaluation measure. (We include \MethodThreeabb-SVM and \MethodThreeabb-\LambdaMARTabb together in this analysis.) We mark the features with (d/p,x): whether the feature is document-based or passage-based (d/p) and the number of cases (x) its removal caused statistically significant performance drop. The ordered list of features is: ESA (p,15), SDM unigrams (d,4), SDM biterms (d,2), SW1 (d,2), Ent (d,1), SW2 (d,1), SDM bigrams (d,1), MaxPDSim (p,1), LengthRatio (p,1), \MKsynonymsOverlap (p,1), pLocation (p,1), Entity (p,1). Thus, as was the case for the SVM-based feature weight analysis from above, ESA which is a passage feature and SDM unigrams which is a document feature are the most important. More generally, the list contains both document and passage features. We note that while the removal of each of the document features resulted in at least one case of statistically significant drop, for quite a few passage features this was not the case; i.e., there is redundancy between the passage features.

We next turn to present feature analysis for the \MethodTwoTitle approach\footnote{In this analysis we set $\rrfParam$, the free parameter of \MethodTwoTitle, to a value which is effective across the train folds.}.
\MethodTwoTitle uses the same document features as \MethodThreeabb, but different passage-based features: mainly those which quantify the rank positions of the document's passages in the passage ranking. The results of an ablation test, as that performed above, are: max (p,5), SW2 (d,4), SDM unigrams (d,3), SDM biterms (d,2), avg (p,2), numPsg (p,2), Ent (d,1), SW1 (d,1), SDM bigrams (d,1), min (p,1), std (p,1), top50 (p,1). We observe again a mix of document and passage features. The max feature, which quantifies the rank position of the document's most highly ranked passage, is more important than the min and avg features. This finding provides further support to the merits of using information about the highest ranked passage of the document.

\begin{table*}[t]
	\tabcolsep=0.1cm
	\caption{\label{tab:JPD_Rankers_Comp} {Varying the LTR method used in JPDs {and in init-LTR}. {'$\sigDocLTR$' marks statisitcally significant difference with init-LTR. Boldface: the best result in a column for each LTR method (\SVMabb, \LambdaMARTabb, \MART or \CAscent)}}}\center \small
	\begin{tabular}{@{}lcccccccccc@{}}
		\toprule
		\multicolumn{1}{l}{}  & \multicolumn{2}{c}{ROBUST} 
		& \multicolumn{2}{c}{WT10G} & \multicolumn{2}{c}{GOV2} 
		& \multicolumn{2}{c}{ClueWeb} & 
		\multicolumn{2}{c}{INEX} \\ \cmidrule(l){2-3} \cmidrule(l){4-5} \cmidrule(l){6-7} \cmidrule(l){8-9} \cmidrule(l){10-11}
		& {\MAP} & {\PTen} & 
		{\MAP} & {\PTen} & 
		{\MAP} & {\PTen} & 
		{\MAP} & {\PTen} & 
		{\MAP} & {\PTen}\\
		
		\midrule
		
		init-\SVMabb  &
		$.261$ & $.439$ &
		$.213$ & $.334$ &
		$.336$ & $.643$ &
		$.222$ & $.406$ & 
		$.392$ & $.577$
		\\
		
		\MethodThreeabb-\SVMabb  &
		
		$\mathbf{.290}^{\sigDocLTR}$ & $\mathbf{.480}^{\sigDocLTR}$ &
		
		$\mathbf{.235}^{\sigDocLTR}$ & $\mathbf{.381}^{\sigDocLTR}$ &
		
		$\mathbf{.350}^{\sigDocLTR}$ & $\mathbf{.656}$ &
		
		$\mathbf{.246}^{\sigDocLTR}$ & $\mathbf{.452}^{\sigDocLTR}$ & 
		
		$\mathbf{.417}^{\sigDocLTR}$ & $\mathbf{.589}$
		
		\\	 \midrule
		
		init-\LambdaMARTabb  &
		$.245$ & $.427$ &
		$.198$ & $.311$ &
		$.326$ & $.651$ &
		$.224$ & $.394$ & 
		$.378$ & $.584$
		\\
		
		\MethodThreeabb-\LambdaMARTabb  &
		
		$\mathbf{.290}^{\sigDocLTR}$ & $\mathbf{.471}^{\sigDocLTR}$ &
		
		$\mathbf{.229}^{\sigDocLTR}$ & $\mathbf{.378}^{\sigDocLTR}$ &
		
		$\mathbf{.345}^{\sigDocLTR}$ & $\mathbf{.655}$ &
		
		$\mathbf{.234}^{\sigDocLTR}$ & $\mathbf{.423}^{\sigDocLTR}$ & 
		
		$\mathbf{.412}^{\sigDocLTR}$ & $\mathbf{.593}$

		\\	\midrule
		
		init-\MART  &
		$.258$ & $.439$ &
		$.203$ & $.305$ &
		$.332$ & $.640$ &
		$.216$ & $.403$ & 
		$.381$ & $.565$
		\\
		
		\MethodThreeabb-\MART  &
		
		$\mathbf{.285}^{\sigDocLTR}$ & $\mathbf{.462}^{\sigDocLTR}$ &
		
		$\mathbf{.211}$ & $\mathbf{.345}^{\sigDocLTR}$ &
		
		$\mathbf{.343}^{\sigDocLTR}$ & $\mathbf{.659}$ &
		
		$\mathbf{.223}$ & $\mathbf{.415}$ & 
		
		$\mathbf{.407}^{\sigDocLTR}$ & $\mathbf{.577}$
		
		\\	\midrule
		
		%
		%
		%
		%
		%
		%
		
		init-\CAscent  &
		$.257$ & $.443$ &
		$.211$ & $.324$ &
		$.329$ & $\mathbf{.649}$ &
		$.212$ & $.406$ & 
		$.377$ & $.586$
		\\
		
		\MethodThreeabb-\CAscent  &
		
		$\mathbf{.273}^{\sigDocLTR}$ & $\mathbf{.471}^{\sigDocLTR}$ &
		
		$\mathbf{.226}^{\sigDocLTR}$ & $\mathbf{.372}^{\sigDocLTR}$ &
		
		$\mathbf{.339}^{\sigDocLTR}$ & $.647$ &
		
		$\mathbf{.215}$ & $\mathbf{.420}$ &
		
		$\mathbf{.382}$ & $\mathbf{.602}$ 
		
		\\	
		
		\bottomrule 
		
	\end{tabular}
	
\end{table*}

{
	\subsubsection{LTR Methods}
	\label{sec:moreLTR}
	Heretofore, we applied our methods using two LTR approaches: \SVM and
	\LambdaMART. In Table \ref{tab:JPD_Rankers_Comp}, we study the
	performance of our \MethodThreeabb method with two additional LTR approaches:
	\MART \citep{friedman:01} and coordinate ascent 
	\citep{metzler+Croft:07}.  \MART, known as gradient boosted regression
	trees, is a non-linear pairwise ranker which combines the outputs
	obtained by different regression trees. On the other hand, coordinate
	ascent (\CAscent in short) is a linear listwise approach.
	We used the RankLib
	implementations of the \MART and \CAscent
	algorithms\footnote{\url{https://sourceforge.net/p/lemur/wiki/RankLib/}.}. \CAscent
	was trained for NDCG@10.}

Table \ref{tab:JPD_Rankers_Comp}
shows that the \MethodThreeabb method improves over the initial LTR ranking in all relevant comparisons (5 datasets $\times$ 2
evaluation measures $\times$ 4 LTR methods). Most of the improvements for SVM and LMart are statistically significant while some of the improvements for MART and CAscent are statistically significant.

We also see in Table \ref{tab:JPD_Rankers_Comp} that in most relevant
comparisons, using \MethodThreeabb with \SVMabb and \LambdaMARTabb
results in performance that transcends that of its implementations that use \MART and \CAscent. This finding
can be attributed to some extent to the effectiveness of the passage
ranking utilized by \MethodThreeabb. The \MAiP effectiveness of the
passage ranking induced using \MART and \CAscent is lower than that attained by using \SVMabb and \LambdaMARTabb when using the INEX dataset for passage retrieval evaluation. Specifically, the \MAiP performance of  \SVMabb, \LambdaMARTabb, \MART and \CAscent is $.267$, $.275$, $.250$ and $.259$, respectively.

\begin{table*}[h]
	\tabcolsep=0.1cm
	\caption{\label{tab:MSLR_JPDs} Using the MSLR (LETOR) document features in comparison to using the features used thusfar for the initial document ranking and in our JPDs method. '$\sigDocLTR$' and `$m$' mark statistically significant differences with \init-LTR and init-\Mslr-LTR, respectively. Boldface: the best result in a column, per block of either the original features (first block) or the MSLR features (second block), for each LTR method (SVM or LMart).}\center
	\small
	\begin{tabular}{@{}lcccc@{}}
		\toprule
		\multicolumn{1}{l}{}  & \multicolumn{2}{c}{GOV2} 
		& \multicolumn{2}{c}{ClueWeb} \\ \cmidrule(l){2-3} \cmidrule(l){4-5}
		
		& {\MAP} & {\PTen} & 
		{\MAP} & {\PTen}\\ 
		\midrule
		
		\init-\SVMabb &
		$.336$ & $.643$ &
		$.222$ & $.406$
		\\

		\init-\LambdaMARTabb  &
		$.326$ & $.651$ &
		$.224$ & $.394$ 
		\\ \midrule

		\MethodThreeabb-\SVMabb  &
		
		$\mathbf{.350}^{\sigDocLTR}$ & $\mathbf{.656}$ &
		
		$\mathbf{.246}^{\sigDocLTR}$ & $\mathbf{.452}^{\sigDocLTR}$ 
		
		\\	
		
		\MethodThreeabb-\LambdaMARTabb  &

		$\mathbf{.345}^{\sigDocLTR}$ & $\mathbf{.655}$ &
		
		$\mathbf{.234}^{\sigDocLTR}$ & $\mathbf{.423}^{\sigDocLTR}$ 
		
		\\	  \midrule

		init-\Mslr-\SVMabb &
		$.323$ & $.595$ &
		$.251$ & $.437$
		\\
		
		init-\Mslr-\LambdaMARTabb  &
		$.315$ & $.599$ &
		$.241$ & $.428$ 
		\\ \midrule
		
		\MethodThreeabb-MSLR-\SVMabb  &
		
		$\mathbf{.353}^{m}$ & $\mathbf{.634}^{m}$ &
		$\mathbf{.264}^{m}$ & $\mathbf{.452}^{m}$
		\\
		
		\MethodThreeabb-MSLR-\LambdaMARTabb  &
		
		$\mathbf{.342}^{m}$ & $\mathbf{.633}^{m}$ &
		$\mathbf{.244}$ & $\mathbf{.437}$ 
		\\ \bottomrule
		
	\end{tabular}
	
\end{table*}

{
	\subsubsection{Using LETOR Features}
	\label{sec:letor}
	Insofar, we have used the document features described in Section
	\ref{sec:docRetEval}. This practice resulted in highly effective
	document ranking performance as exhibited by the init-LTR baselines
	as well as our methods. We now turn to explore the performance of our methods with a much larger set of document(-query) features. Specifically, we use the
	\Mslr\footnote{\url{www.research.microsoft.com/en-us/projects/mslr}}
	features from the LETOR datasets for retrieval over the GOV2 and ClueWeb collections with the queries specifed in Table \ref{tab:datasets}. We used all MSLR features except for the Outlink number, SiteRank,
	QualityScore, QualityScore2, Query-url click count, url click count,
	and url dwell time. In addition to the MSLR features, we also use here
	the highly effective query-independent document quality measures used above:
	the fraction of terms in the document that are stopwords,
	the fraction of stopwords that appear in the document, and the
	entropy of the term distribution in the document. The stopword list
	used for the two stopword features is composed of the collection's 100
	most frequent alphanumeric terms \citep{ntoulas+al:06,raifer+al:17}.
	For ClueWeb we also used the spam score assigned to a
	document by the Waterloo spam classifier and the PageRank score. All together, we used, at the document level, $149$ features for GOV2 and $151$ features for ClueWeb. }

{The results are presented in Table
	\ref{tab:MSLR_JPDs}. We first see that in terms of the initial ranking, the MSLR features are more effective than those we used above for ClueWeb, but the reverse holds for GOV2. (This could potentially be attributed to the fact that for GOV2 there are fewer queries than for ClueWeb.) We further see in Table \ref{tab:MSLR_JPDs} that our JPDs method is also effective with the MSLR features. It always outperforms the initial ranking; in most relevant comparisons, the improvements are statistically significant.
}

\begin{table*}[t]
	\tabcolsep=0.1cm
	\small
	\caption{\label{tab:psgRetrievalMain_INEX} Passage retrieval over INEX with passages of length 300, 150 and 50. LM is standard language-model-based document retrieval (i.e., documents serve for passages). 
		Boldface: the best result in a column. Statistically
		significant differences with \QL, \QuerySimFuse 
		and \PLMShort are marked with '$\sigDoc$', '$\sigQSF$' and '$\sigPLM$',
		respectively. '$\sigOwpc$' and '$\sigMKS$' mark a statistically
		significant difference between \psgLTR-X and \owpc-X and between \psgLTR-X
		and \MKS-X, respectively. }
	\center
	\begin{tabular}{@{}lccccccccc@{}}	
		\toprule
		\multicolumn{1}{l}{} & \multicolumn{9}{c}{\inex} \\ \cmidrule(l){2-10}
		\multicolumn{1}{l}{} & \multicolumn{3}{c}{\psgLength{300}} & \multicolumn{3}{c}{\psgLength{150}} & \multicolumn{3}{c}{\psgLength{50}} 
		\\ \cmidrule(l){2-4} \cmidrule(l){5-7} \cmidrule(l){8-10}
		
		\multicolumn{1}{l}{} & {\MAiP} & {\officialiP} & {\officialiPOne} & 
		{\MAiP} & {\officialiP} & {\officialiPOne} & 
		{\MAiP} & {\officialiP} & {\officialiPOne} \\
		
		\midrule
		
		\QL & $.256$ & $.523$ & $.449$ & 
		$.256$ & $.523$ & $.449$ & 
		$.256$ & $.523$ & $.449$ \\
		
		
		
		\QuerySimFuse & 
		$.248$ & $.577$ & $.453$ & 
		$.234$ & $.575$ & $.455$ &
		$.209$ & $.581$ & $.449$ \\
		
		\PLMShort & 
		$.253$ & $.586$ & $.472$ &
		
		$.240$ & $.596$ & $.471$ & 
		
		$\mathbf{.215}$ & $.605$ & $\mathbf{.469}$ \\
		
		\midrule
		
		\owpc-\SVMabb & $.242$ & $.577$ & $.440$ & 
		$.229$ & $.569$ & $.438$ &
		$.202$ & $.570$ & $.431$ \\
		
		\owpc-\LambdaMARTabb & $.255$ & $.578$ & $.460$ &
		$.240$ & $.566$ & $.450$ &
		$.208$ & $.577$ & $.443$ \\
		
		\midrule
		
		\MKS-\SVMabb & $.247$ & $.593$ & $.468$ &
		$.235$ & $.602$ & $.459$ &
		$.199$ & $.626$ & $.457$ \\
		
		\MKS-\LambdaMARTabb & $.262$ & $.620$ & $.479$ &
		$.241$ & $.629$ & $.479$ &
		$.200$ & $.644$ & $.459$ \\
		
		\midrule
		
		\psgLTR-\SVMabb & 
		$.267_{\sigOwpc\sigMKS}$ &  $.637^{\sigDoc\sigQSF}_{\sigOwpc}$ & $.487_{\sigOwpc}$ &
		
		$\mathbf{.253_{\sigOwpc}}$ & $\mathbf{.662^{\sigDoc\sigQSF\sigPLM}_{\sigOwpc\sigMKS}}$ & $.492_{\sigOwpc}$ &
		
		$.213^{\sigDoc}$ & $\mathbf{.647^{\sigDoc}_{\sigOwpc}}$ &  $.467$ \\
		
		\psgLTR-\LambdaMARTabb & 
		$\mathbf{.275^{\sigQSF\sigPLM}_{\sigOwpc}}$ & $\mathbf{.644^{\sigDoc\sigQSF\sigPLM}_{\sigOwpc}}$ & $\mathbf{.496}$ &
		
		$\mathbf{.253}$ & $.650^{\sigDoc\sigQSF}_{\sigOwpc}$ & $\mathbf{.494_{\sigOwpc}}$ &
		
		$.209^{\sigDoc}$ & $.634^{\sigDoc}$ & $.454$ \\
		
		\bottomrule
	\end{tabular}
	 
\end{table*}

\begin{table}[h]
	\tabcolsep=0.1cm
	
	\caption{\label{tab:senRetrievalMain_AQUAINT} Sentence retrieval over AQUAINT. Boldface: the best result in a column. 
		Statistically significant differences with \QuerySimFuse
		and \PLMShort are marked with '$\sigQSF$' and '$\sigPLM$',
		respectively. '$\sigOwpc$' and '$\sigMKS$' mark a statistically
		significant difference between \psgLTR-X and \owpc-X and between \psgLTR-X and \MKS-X, respectively.  }
	\center
	\small
\begin{tabular}{@{}lcc@{}}	
	\toprule
	
	\multicolumn{1}{l}{} & \multicolumn{2}{c}{\aquaint} \\ \cmidrule(l){2-3}
	\multicolumn{1}{l}{} & 
	{\MAP} & {\PTen} \\
	
	\midrule 
	
	\QuerySimFuse & $.471$ & $.624$
	\\
	\PLMShort & $.518$ & $.669$
	\\
	\midrule
	
	\owpc-\SVMabb & $.579$ & $.701$
	\\
	
	\owpc-\LambdaMARTabb & $.589$ & $\mathbf{.716}$
	\\
	\midrule
	\MKS-\SVMabb & $.569$ & $.664$
	\\
	
	\MKS-\LambdaMARTabb & $.585$ & $.701$
	\\
	\midrule
	\psgLTR-\SVMabb & $.602^{\sigQSF\sigPLM}_{\sigOwpc\sigMKS}$ & $.713^{\sigQSF}_{\sigMKS}$
	\\
	\psgLTR-\LambdaMARTabb & $\mathbf{.606^{\sigQSF\sigPLM}_{\sigOwpc\sigMKS}}$ & $.710^{\sigQSF}$
	\\
	\bottomrule
	
\end{tabular}

\end{table}

\subsection{Passage Retrieval}
\label{sec:psgRetResults}
Heretofore, we have focused on the document retrieval task. Our
passage-based document retrieval methods utilize a ranking of passages induced using our
proposed passage retrieval approach. (See Section \ref{sec:pRetF} for
details.) We now turn to compare the performance of our passage ranker with that of the passage retrieval baselines described in Section \ref{sec:psgRetEval}.

Table \ref{tab:psgRetrievalMain_INEX} presents the performance numbers of
the passage retrieval methods for the \inex collection. 
We see that our LTR methods, \psgLTR-\SVMabb
and \psgLTR-\LambdaMARTabb, outperform all other passage retrieval methods
in most relevant comparisons (3 passage lengths $\times$ 3 evaluation measures) 
with many of the improvements being
statistically significant. We note that the \MKS baseline \citep{yang+al:2016} was recently shown to yield state-of-the-art passage retrieval performance.

Table \ref{tab:senRetrievalMain_AQUAINT} presents the effectiveness of our passage retrieval approach, \psgLTR, in ranking sentences in the \aquaint collection.
We see that \psgLTR-\SVMabb and \psgLTR--\LambdaMARTabb statistically significantly outperform all other passage retrieval methods in terms of \MAP. In the single case where our methods are outperformed by another method (MKS-LMart) in terms of p@10, the performance differences are not statistically significant.

The findings presented above for focused (passage) retrieval over INEX, and sentence retrieval over AQUAINT, attest to the fact that our passage ranker posts state-of-the-art passage retrieval performance.

\subsubsection{Feature Analysis for Passage Retrieval}
We first use the SVM-based feature analysis, as was performed above for document retrieval, to analyze the relative importance of features used in our passage retrieval approach (\psgLTR-SVM). For INEX, we consider each of the three passage lengths as a different experimental setting.
The top $10$ features for \inex are: \ESA, \swOne, \pMax,
\EntityLinking, \std, \swTwo, \entropy, \documentSim, \pAvg and
\MKsynonymsOverlap.  For \aquaint, the top-10 features are: \entropy,
\swOne, \ESA, \lengthRatio, \MKtermOverlap, \pAvg, \nighborLeft,
\MKsynonymsOverlap, \nighborRight, \MKgLength. Recall that using
stopwords-based passage priors (\swOne and \swTwo) to rank passages is
novel to this study.  We see that \swOne is the second most important
feature for both \inex and \aquaint. Another observation is that, as
expected, the relative ordering of passages in this analysis, and the
set of features that are among the top-10, are not identical to those
presented above when using the passage features for document
retrieval.

In addition, we perform ablation tests for \psgLTR. When using passages of 300 terms for INEX, the features whose removal resulted in statistically significant performance drop of MAiP are: ESA, MaxPDSim, AvgPDSim, SW1. The features whose removal resulted in statistically significant performance drop of MAP for AQUAINT are: Ent, SW1, ESA, SW2. The features are ordered in both cases in a descending order of the performance drop. Given that the retrieval tasks over INEX (passage retrieval) and AQUAINT (sentence retrieval) are different, it is not surprising that the feature lists are a bit different. Yet, ESA and SW1 are in both cases among the most important features, which was also the case above in the SVM-based analysis.

\section{Conclusions and Future Work}
Our focus in this work was on passage-based document retrieval:
document ranking methods that utilize information induced from
document passages.  Previous work on passage-based document retrieval
has focused on methods that integrate passage-query and
document-query similarity values. Here, we addressed the challenge of
utilizing richer sources of passage-based information for improving
document retrieval effectiveness.

We presented a suite of learning-to-rank methods for document retrieval that use passage-based information. Most of the methods rely on ranking passages in response to the query using an effective approach, specifically, utilizing learning-to-rank. Some of the methods use information about the ranking of the passages of a document. Other methods use the passage-based features utilized for passage ranking and integrate them with document-based features so as to learn a document ranking function. We described connections between our methods and past unsupervised approaches for passage-based document retrieval as well as approaches for ranking clusters of similar documents.

To learn a passage-ranking method, we used previously proposed features along with features which were not used before for learning passage ranking functions. These  features are query-independent passage-relevance priors adopted from work on using document relevance priors for Web search.

Empirical evaluation performed with a suite of datasets demonstrated the effectiveness of our methods. Our most effective method integrates document-based features with passage-based features of the document's most highly ranked passage. 
{In addition, our best performing method was shown to outperform the use of different sets of document-based features}. Further exploration provided support to the merits of using an effective passage ranking method. We also showed that our passage-ranking method yields state-of-the-art passage retrieval performance.

For future work we intend to integrate in our methods additional passage-based
features; e.g., those induced from inter-passage similarities. We also plan to explore how our methods can be used for, and with, pseudo-feedback-query expansion. A case in point, we can apply query expansion at the passage-level, document-level, or both, so as to enrich the feature set used.

\end{document}